\documentclass[fleqn,usenatbib]{mnras}
\usepackage{newtxtext,newtxmath}
\usepackage[T1]{fontenc}
\usepackage{ae,aecompl}
\usepackage{graphicx}	% Including figure files
\usepackage{amsmath}	% Advanced maths commands
\usepackage{amssymb}	

\usepackage{bigints}    %larger integrals
\usepackage{bm}
\usepackage{algorithmic}
\usepackage{algorithm}
\usepackage{units}
\usepackage{amsmath}
\usepackage{graphicx}
\usepackage{tabularx}
\usepackage{cleveref}

\crefformat{section}{\S#2#1#3} % see manual of cleveref, section 8.2.1
\crefformat{subsection}{\S#2#1#3}
\crefformat{subsubsection}{\S#2#1#3}
\usepackage{ulem}
\usepackage[dvipsnames]{xcolor}

\newcommand{\be}{\begin{equation}}
\newcommand{\ee}{\end{equation}}
\newcommand{\ba}{\begin{eqnarray}}
\newcommand{\ea}{\end{eqnarray}}

\newcommand{\vc}{\bm{c}}
\newcommand{\va}{\bm{a}(z)}
\newcommand{\vb}{\bm{b}(z)}
\newcommand{\vCo}{\bm{C}_{\rm obs}}

\newcommand{\vCi}{\bm{C}_{\rm int}(z)}

\newcommand{\vCt}{\bm{C}_{\rm tot}(z)}
\newcommand{\vCti}{\bm{C}_{\rm tot}^{-1}(z)}

\newcommand{\mrefz}{m_{i, \mathrm{ref}}}
\newcommand{\mi}{m_{i}}
\newcommand{\dense}{\mathtt{dense}}
\newcommand{\lum}{\mathtt{luminous}}
\newcommand{\dk}{\boldsymbol{\gamma}_{\alpha}^{(k)}}
\newcommand{\dg}{\boldsymbol{\gamma}_{t}}
\newcommand{\dbar}{\overline{\boldsymbol{\gamma}_{\alpha}}}
\newcommand{\njk}{N_{\rm JK}}

\title[KiDS LRGs]{Luminous red galaxies in the Kilo Degree Survey: selection with broad-band photometry and weak lensing measurements}

\author[M. Vakili et al.]{
Mohammadjavad Vakili$^{1}$\thanks{E-mail: vakili@mail.strw.leidenuniv.nl},
Maciej Bilicki$^{1,2}$,
Henk Hoekstra$^{1}$,
Nora Elisa Chisari$^{3}$,
\newauthor
Christos Georgiou$^{1}$,
Arun Kannawadi$^{1}$,
Koen Kuijken$^{1}$,
Angus  H.  Wright$^{4}$
\\
$^{1}$Leiden Observatory, Leiden University, Leiden, Netherlands\\
$^{2}$National Centre for Nuclear Research, Astrophysics Division, P.O. Box 447, 90-950 \L{}\'{o}d\'{z}, Poland\\
$^{3}$Department of Physics, University of Oxford, Keble Road, Oxford, OX1 3RH, UK\\
$^{4}$Argelander-Institut f\"ur Astronomie, Auf dem H\"ugel 71, 53121 Bonn, Germany
}

\date{Accepted XXX. Received YYY; in original form ZZZ}

\pubyear{2018}

\begin{document}
\label{firstpage}
\pagerange{\pageref{firstpage}--\pageref{lastpage}}
\maketitle

\begin{abstract}

We use the overlap between multiband photometry of the Kilo-Degree Survey (KiDS) and spectroscopic data based on the Sloan Digital Sky Survey (SDSS) and Galaxy And Mass Assembly (GAMA) to infer the colour-magnitude relation of red-sequence galaxies. We then use this inferred relation to select luminous red galaxies (LRGs) in the redshift range of $0.1<z<0.7$ over the entire KiDS Data Release 3 footprint. We construct two samples of galaxies with different constant comoving densities and different luminosity thresholds. The selected red galaxies have photometric redshifts with typical photo-z errors of $\sigma_z \sim 0.014 (1+z)$ that are nearly uniform with respect to observational systematics. This makes them an ideal set of galaxies for lensing and clustering studies. As an example, we use the KiDS-450 cosmic shear catalogue to measure the mean tangential shear signal around the selected LRGs. We detect a significant weak lensing signal for lenses out to $z \sim 0.7$. 

\end{abstract}

%\textbf{}\clearpage
%%%%%%%%%%%%%%%%%%%%%%%%%%%%%%%%%%%%%%%%%%%%%%%%%%

%%%%%%%%%%%%%%%%% BODY OF PAPER %%%%%%%%%%%%%%%%%%

\section{Introduction}

The Kilo Degree Survey (KiDS) is a wide-angle optical survey designed, among others, to map the dark matter distribution by studying the weak gravitational lensing of galaxies (\citealt{kuijken2015}). This is done by measuring the correlation between the distortion of the shapes of distant galaxies. These correlations are then compared to the predictions of 
cosmological simulations to test cosmological models (\citealt{heymans2013,jee2016,hendrick2017,joudaki2017,troxel2017}). 

However, the full constraining power of weak lensing studies can be unlocked through joint 
analysis of the cosmic shear of background galaxies (known as source galaxies) and the positions of foreground lens galaxies that have robust distance estimates -- either from spectroscopic or precise and accurate photometric redshifts. This procedure, known as galaxy-galaxy lensing, can be used for tightening the lensing constraints 
on cosmological parameters (see \citealt{cacciato2013,elvin2017,joudaki2018,edo2018}) by mitigating the biases arising from observational and astrophysical 
systematics. Furthermore, it helps us understand the connection between the properties of the foreground galaxies and the properties of the dark matter halos hosting them (\citealt{viola2015,edo2016,clampitt2017,dvornik2018}).

Furthermore, measurements of the intrinsic alignments of galaxies (see \citealt{hirata2004,kirk2015} and references therein) can benefit from having a sample of galaxies with known redshifts (\citealt{mandelbaum2011,singh2015,fastsound2017}) or photometric redshifts with small uncertainties (\citealt{joachimi2009,joachimi2010,joachimi2011}). Another application of a galaxy sample with robust redshifts is the calibration of the photometric redshift distributions of source galaxies in weak lensing surveys using cross correlation of the two samples (\citealt{cawthon2017,davis2017,hendrick2017,morrison2017}).

In weak lensing surveys, photometric redshifts are often obtained by template fitting or machine learning techniques. Redshifts derived from the former method are based on the assumption that galaxy fluxes computed from multi-band photometry can be expressed as a superposition of a set of templates and some prior over the types of galaxies \citep[e.g.][]{bpz1999,Bolzonella2000,Feldmann2006,Brammer2008}. Machine learning methods make use of the overlap between the imaging surveys and spectroscopic data to find the complex relation between galaxy colours and their redshifts \citep[e.g.][]{Firth2003,Wadadekar2005,Way2009,Gerdes2010}. Additionally, hybrid approaches joining template fitting with machine learning are being investigated \citep[e.g.][]{boris2017,Duncan2018}.

% On the other hand, \citet{boris2017} showed that it is possible to formulate a hybrid approach in which a spectroscopic training sample (albeit incomplete and heterogeneous) can be used to infer the set of templates and the redshifts of galaxies simultaneously. \MB{[I think we can organize it a bit better: some key canonical references for template-fitting, ML, and hybrid approaches. I'll look into that later]}

An alternative way to derive robust redshifts is by taking advantage of the properties of galaxies with old stellar populations. Such objects can be efficiently selected from multi-band photometry of imaging surveys without the need of full spectroscopic coverage for each single source. At any given redshift, the distribution of these galaxies in the colour-magnitude diagram follows a straight line --- with some intrinsic scatter --- known as the red-sequence ridge-line. Therefore, these galaxies are called the red-sequence galaxies. The distribution of the red-sequence galaxies in the colour-magnitude diagram permits us to separate these galaxies from the rest of the galaxy population (\citealt{gladders_yee2000,hao2009,redmap_sdss,rozo2016}). 

For a sample of red-sequence galaxies with spectroscopic redshifts, one can parametrize the redshift evolution of the red-sequence ridge-line, also known as the red-sequence template. Assuming a prior probability over the redshifts of red galaxies and a redshift-dependent distribution over the magnitudes of red galaxies, the red-sequence template can be turned into a red-sequence selection algorithm in photometric data. Furthermore, the redshifts of the selected galaxies can be precisely estimated without obtaining spectroscopy for them. This procedure, known as $\textsc{redMagiC}$, has been successfully applied to the Sloan Digital Sky Survey and the Dark Energy Survey data (\citealt{rozo2016}). Obtaining a sample of galaxies with a well-defined selection and precise redshifts over the entire footprint of a given galaxy survey has been proven beneficial for galaxy-galaxy lensing studies (\citealt{clampitt2017,prat2017}), galaxy clustering (\citealt{elvin2017}), and joint cosmological probes. 

In this investigation, we select a set of red-sequence galaxies from the overlap of the KiDS DR3 (\citealt{kids_dr3}) multi-band photometry and the spectroscopic redshift surveys of SDSS and GAMA. These galaxies are then used to calibrate the red-sequence template. We then follow the $\textsc{redMagiC}$ prescription (\citealt{rozo2016}) to select the red-sequence galaxies and estimate their redshifts. After imposing a set of luminosity cuts and constant comoving densities, we construct two samples of luminous red galaxies suitable for cross-correlation studies.

We then compare the derived red-sequence redshifts of the selected galaxies in this work with the photometric redshifts derived from other methods. Based on overlapping spectroscopy from SDSS, GAMA, as well as 2dFLenS \citep{blake2016}, we investigate the dependence of the photo-$z$ errors on the variation of observational systematics across the survey tiles. Using the KiDS-450 cosmic shear data (\citealt{lensfit4,hendrick2017}), we present measurement of the weak lensing signal using the red galaxies as lenses and we find significant detection of the mean tangential shear signal. Finally, we investigate if the weak lensing measurements can pass a set of systematic null tests. The main purpose of this work is to present a sample of photometrically selected LRGs with robust redshifts. The lensing measurements are presented as a straightforward use case of the sample. However, the applications and modelling of the clustering and lensing of this sample are left for future work.
%For instance, joint analysis of cosmic shear signal 
%obtained from faint background galaxies and clustering of foreground galaxies can be utilized to break the degeneracy between constraints on cosmological parameters and to gain insights into 
%how galaxies trace the underlaying dark matter structure. 

The structure of the paper is as follows. 
The characteristics of the datasets, both photometric and spectroscopic, are described in Section~\ref{sec:data}. 
In Section~\ref{sec:methodology} we introduce the methodology used in this analysis including the selection of seed red-sequence galaxies (red-sequence galaxies with secure spectroscopic redshifts for estimating the colour-magnitude relation), inference of the red-sequence colour magnitude relation, and selection of the final LRG sample based on appropriate cuts on the estimated luminosities and the quality of red-sequence fits. 
In Section~\ref{sec:photoz} we describe the two samples of LRG candidates identified by applying two luminosity ratio thresholds and by imposing two constant comoving number densities. We discuss the photometric redshift performance of the selected red galaxy catalogues by comparing the derived red-sequence redshifts with spectroscopic redshifts. Furthermore, we compare the red-sequence redshifts estimated in this work with other photo-$z$ solutions available in KiDS DR3. We also discuss the impact of observing conditions on the estimated LRG red-sequence photo-$z$'s. We then present the weak lensing measurements and a set of lensing systematic tests in Section~\ref{sec:lensing}. Finally, we summarize and conclude in Section~\ref{sec:summary}. 

Note that calculating the comoving densities and distances requires specifying a cosmology. In this work, we assume a flat $\Lambda$CDM cosmology with $\Omega_{m} = 0.3$ and $h=1.0$ \footnote{This is the convention used by \citet{redmap_sdss} in constructing the SDSS redMaPPer catalogue.}. All distances and comoving densities are quoted in units of $h^{-1}\; \mathrm{Mpc}$ and $h^{3} \; \mathrm{Mpc}^{-3}$ respectively. Also note that the luminosity ratios used for selection of the red galaxies are not sensitive to the choice of $h$ and in this work and we always work with luminosity ratios. Whenever magnitudes are used, they will be provided in the AB system.

\section{Data}\label{sec:data}

\subsection{KiDS photometric data}\label{sec:kids}
The Kilo-degree Survey (KiDS, \citealt{kids}) is a wide 
imaging survey conducted with the OmegaCAM camera (\citealt{omegacam}) which is mounted on the VLT Survey Telescope (\citealt{vst}). This survey uses four broad-band filters ($ugri$) in the optical wavelengths. KiDS targets approximately 1350 deg$^2$ of the sky in two regions, one on the celestial equator and the other one in the South Galactic cap.

The latest public data release of KiDS is the third data release (DR3, \citealt{kids_dr3}) which covers $\sim 450$ deg$^{2}$ of the sky with 5$\sigma$ depth of 24.3, 25.1, 24.9, 23.8 in $2$ arcsec apertures in the $ugri$ bands respectively. For a thorough description of the KiDS data reduction, we refer the readers to the data release paper (\citealt{kids_dr3}). 

The KiDS database includes magnitudes derived by SExtractor (\citealt{sextractor}) such as \texttt{ISO} and \texttt{AUTO}. These magnitudes are determined directly from images with a variety of PSF values, they are therefore not optimal for our purposes where colours independent of such variations are needed. The KiDS data reduction involves however a post-processing procedure in which Gaussian Aperture and PSF (GAaP,~\citealt{gaap}) magnitudes are derived (\citealt{kuijken2015}). This procedure is performed in the following way. First, the PSF is homogenized across each individual coadd. Afterwards, a Gaussian-weighted aperture is used to measure the photometry. The size and shape of the aperture is determined by the length of the major axis, the length of the minor axis, and the orientation, all measured in the $r$-band. This procedure provides a set of magnitudes for all filters. 

The magnitudes used in this work are the zeropoint-calibrated and extinction-corrected magnitudes\footnote{In the final catalogue and for each band, the zeropoint offsets ($\mathtt{ZPT}_{-}\mathtt{offset}_{-}\mathtt{band}$) and the Galactic extinction corrections ($\mathtt{EXT}_{-}\mathtt{SFD}_{-}\mathtt{band}$) based on \citet{schlegel98} are provided in separate columns.} 
denoted by $\; \mathtt{Mag}_{-}\mathtt{type}_{-}\mathtt{band}_{-}\mathtt{calib}$. The default magnitudes in KiDS are GAaP magnitudes. They were designed to provide accurate colours but underestimate total fluxes of large galaxies. Total fluxes are, however, needed in our LRG selection procedure to derive luminosities (see section~\ref{sec:overview}). Therefore, whenever galaxy fluxes are needed, we use $\mathtt{Mag}_{-}\mathtt{AUTO}_{-}\mathtt{band}$ in our red-sequence modelling. 

For our choice of colour, GAaP colours are used as they have less scatter and bias than the colours derived from the $\mathtt{Mag}_{-}\mathtt{AUTO}$ magnitudes. For the rest of this paper, we work with the calibrated $\mathtt{AUTO}$ magnitudes and GAaP colours and we refer the readers to \citet{kuijken2015} and \citet{kids_dr3} for a more detailed discussion of the derivation of GAaP colours.

%Zeropoints are adjusted by first homogenizing the photometry in the $r$ and $u$ bands using the coadd overlaps in these two filters. Afterwards, the $g$ and the $i$ bands are tied to the $r$ band using stellar locus regression, which homogenizes the $g − r$ and $r − i$ colours.

%The photometric homogenization is done using the GAaP photometry, and in the final catalogues the resulting zeropoint offsets for each band (denoted by $\mathtt{ZPT}_{-}\mathtt{offset}_{-}\mathtt{band}$) are reported
%in separate columns, together with Galactic extinction corrections (denoted by $\mathtt{EXT}_{-}\mathtt{SFD}_{-}\mathtt{band}$) which are based on the \citet{schlegel98} maps. The zeropoint-calibrated and extinction-corrected magnitudes are denoted as $\; \mathtt{Mag}_{-}\mathtt{type}_{-}\mathtt{band}_{-}\mathtt{calib}$:

%\begin{eqnarray}
%\mathtt{Mag}_{-}\mathtt{type}_{-}\mathtt{band}_{-}\mathtt{calib} =  \nonumber\\
%\mathtt{Mag}_{-}\mathtt{type}_{-}\mathtt{band} + \mathtt{ZPT}_{-}\mathtt{offset}_{-}\mathtt{band} - \mathtt{EXT}_{-}\mathtt{SFD}_{-}\mathtt{band}, 
%\end{eqnarray}
%where the uncalibrated magnitude measurements (denoted by $\mathtt{Mag}_{-}\mathtt{type}_{-}\mathtt{band}$) are directly taken from the KiDS multiband catalogue. 

The photometric catalogue is cleaned by removing the artefacts corresponding to any of the following masking flags: readout spike, saturation core, diffraction spike, secondary halo, or bad pixels. Furthermore, only objects for which photometric errors in all bands are provided, are kept in the final photometric catalogue (see \citet{kids_dr3} and \citet{radovich2017}). Finally, we require the final sample to not contain point-like objects by applying the cut $\mathtt{SG2DPHOT}=0$. This parameter is a KiDS star/galaxy classifier based on the $r$ band morphology, and it is equal to 0 for objects that are classified as galaxies.

\subsection{Spectroscopic data}\label{sec:spec}

In this work, we exploit the overlap between the KiDS catalogue and a number of spectroscopic datasets for two purposes. First, we need a set of galaxies in the KiDS catalogue with spectroscopic redshifts that can be used as seeds for estimating the parameters of the red-sequence template. This procedure is explained in detail in section~\ref{sec:seed} and it is applied to the overlap between the KiDS photometry and spectroscopic catalogues of galaxies in GAMA (\citealt{driver2011}) and SDSS DR13 (\citealt{sdss_dr13}). Later, for testing the performance of the redshifts estimated for the selected LRGs in section~\ref{sec:performance} we make use of the overlap between KiDS and the spectroscopic redshifts from SDSS, GAMA, as well as 2dFLenS (\citealt{blake2016}).
In what follows in the rest of this section, we provide a brief description of these spectroscopic catalogues.

\subsubsection{GAMA}
Galaxy And Mass Assembly (GAMA,~\citealt{driver2011}) is a spectroscopic survey  which used the AAOmega spectrograph mounted on the Anglo-Australian Telescope. This survey spans five fields: G09, G12 and G15 on the celestial equators, and G02 and G23 on the Southern Galactic Cap. The only GAMA field outside the KiDS DR3 footprint is G02. The magnitude limited sample of GAMA is nearly complete down to $r=19.8$ mag for galaxies in the equatorial fields and down to $i=19.2$ mag for galaxies in the G23 region (\citealt{likse2015}). The GAMA spectra in the four fields that overlap with KiDS amount to a total of $\sim 230,000$ KiDS sources with high-quality spectroscopic redshifts with $\langle z \rangle = 0.23$. 

\subsubsection{SDSS}

The Sloan Digital Sky Survey (SDSS, \citealt{york2000}) is a photometric and spectroscopic survey of $14,555$ deg$^2$ of the sky encompassing more than one third of the celestial sphere using a dedicated 2.5-m telescope (\citealt{gunn2006}). In particular, we make use of the spectroscopic dataset from the Data Release 13 (DR13, \citealt{sdss_dr13}) of the SDSS-IV project. We only use sources with class `GALAXY'. 

The overlap between SDSS and KiDS in the equatorial fields above $\delta =-3$ gives us $\sim 57,000$ SDSS spectroscopic galaxies with KiDS photometry. However those with $r<19.8$ are mostly included in GAMA, and after removing the latter we are left with nearly $43,000$ unique SDSS spectroscopic galaxies with KiDS photometry. 

The SDSS-matched KiDS galaxies (after removing the overlap with GAMA) span higher redshifts than the GAMA-matched KiDS sources. Furthermore, this sample of galaxies mostly encompasses LRGs that are observed in the Baryonic Oscillation Spectroscopic Survey (BOSS, \citealt{dawson2013}) and the extended BOSS (eBOSS, \citealt{dawson2016}). This makes them ideal candidates for seed galaxies needed to estimate the red-sequence template as we seek to select galaxies that populate the same volume in the colour space as the SDSS LRGs do. 

\subsubsection{2dFLenS}
The 2-degree Field Lensing Survey (2dFLenS, \citealt{blake2016}) is a spectroscopic survey performed at the Australian Astronomical Observatory covering an area of 731 deg$^2$. By expanding the overlap with the KiDS field in the southern galactic cap, this survey aims to provide a dataset suitable for joint clustering and lensing analyses (\citealt{amon2017,joudaki2018}), photometric redshift calibration (\citealt{johnson2017,wolf2017,kids_annz}), and lensing systematic tests (\citealt{amon2018}).

In KiDS DR3 there are nearly $12,000$ galaxies with 2dFLenS spectra. After excluding the galaxies in common with GAMA and SDSS, we have approximately $9,000$ unique 2dFLenS galaxies with KiDS photometry.

\section{Methodology}\label{sec:methodology}

\subsection{Algorithm overview}\label{sec:overview}

At any given redshift, red-sequence galaxies follow a narrow ridge-line in the colour magnitude space. As detailed in \citet{rozo2016}, the reference band used for describing the colour-magnitude relation should lie redwards of the 4000 \AA\ break at all considered redshifts, therefore it is preferable to choose the magnitude of the reddest available bandpass for this. In the KiDS imaging data, the colour vector $\mathbf{c}$ corresponds to the GAaP colours $\{u-g,g-r,r-i\}$ and the magnitude of the reddest photometric bandpass corresponds to to the apparent $i$-band magnitude $m_i$ (\citealt{kids_dr3}). 

This red-sequence colour magnitude relation, also known as the red-sequence template, can be used to characterize the probability distribution function 
$p(\vc|m_i,z)$. This is the probability that a given galaxy with apparent $i$-band magnitude $m_i$ and redshift $z$ has a certain multi-dimensional colour vector $\vc$. At a given redshift $z$, the expected value of $\vc$ is given by a straight line in the space of $\{m,\vc\}$. We denote the redshift and magnitude-dependent expected value of $\vc$ by $\vc_{\rm red}(m_i, z)$:
\be
\vc_{\rm red}(m_i, z) = \langle c|m_i,z\rangle = \int \mathrm{d}\vc \; \vc p(\vc|m_i,z).   
\ee

Since $\vc_{\rm red}(m_i, z)$ is linearly dependent on $m_i$, the relation between $\vc_{\rm red}(m_i, z)$ and $m_i$ can be fully determined by the following parameters: the intercept of the colour-magnitude ridge-line $\va$, the slope of the ridge-line $\vb$, and the reference apparent $i$-band magnitude $\mrefz(z)$\footnote{The choice of $\mrefz(z)$ is arbitrary and it is selected by the investigator. In the next section we will explain how this parameter is set in our analysis.}: 

\be
\vc_{m_i,\rm red}(z) = \va + \vb \big(\mi - \mrefz(z)\big) \label{eq:temp1}
\ee

Moreover, for every galaxy in the survey, we can define a total colour covariance matrix $\vCt$. This matrix is composed of two components: the observed colour covariance $\vCo$ and the intrinsic red-sequence colour covariance $\vCi$:
\be
\vCt = \vCo +\vCi \label{eq:temp2}
\ee

Finally, we assume that the conditional probability density $p(\vc|m_i,z)$ is a multivariate Gaussian with the mean $\vc_{\rm red}(m_i,z)$ given by Eq.~\ref{eq:temp1} and the covariance $\vCt$ given by Eq.~\ref{eq:temp2}. Therefore $p(\vc|m_i,z)$ can be written as:

\be
p(\vc|m_i,z) = \mathcal{N}(\vc \; ;\; \vc_{\rm red}(m_i,z)\;,\; \vCt). \label{eq:temp}
\ee
As we will see later, it is convenient to define a red-sequence chi-squared $\chi^{2}_{\rm red}$:
\be
\chi^{2}_{\rm red} = \big(\vc  - \vc_{\rm red}(z,m_i)\big)^{\rm T} \vCti \big(\vc  - \vc_{\rm red}(z,m_i)\big),
\ee
which is related to $p(\vc|m_i,z)$ in the following way: 
\be
-2 \; \ln \; p(\vc|m_i,z) = \chi^{2}_{\rm red} + \ln \; \Big((2\pi)^{3}\mathrm{det}\big(\vCt\big)\Big). \label{eq:temp_chi}
\ee

Thus, in order to determine the colour-magnitude relation, we are required to estimate the three-dimensional (3D) vectors $\va$, $\vb$, the scalar $\mrefz(z)$, and the 3$\times$3 intrinsic covariance matrix $\vCi$. Hereafter in this work, we ignore the off-diagonal elements of the intrinsic covariance matrix as we expect the intrinsic scatter of red-sequence galaxies to be smaller than the observed photometric uncertainties.

%they are sub-dominant to the off-diagonal elements of the observed colour covariance and the rest of the components of the colour covariance matrix.   

With the red-sequence colour-magnitude relation, $p(\vc|m_i,z)$, at hand, one can estimate the redshift probability distribution function of a galaxy conditioned on the 3D colour vector $\vc$ and the $i$-band magnitude $\mi$. According to Bayes' rule, this probability distribution is given by  

\be
p(z|\mi,\vc) \propto p(\vc|\mi,z)p(\mi|z)p(z),
\label{eq:pzmc}
\ee
Note that in addition to $p(\vc|m_i,z)$ which we have discussed thus far, there are two probability distributions on the right hand side of Eq.~\ref{eq:pzmc}: the distribution of the $i$-band magnitudes of red galaxies $p(\mi|z)$, and the prior distribution over the redshifts of red-sequence galaxies, $p(z)$. 

The magnitude distribution acts as a redshift-dependent luminosity filter and its functional form is assumed as the \cite{schecter1976} function:

\be 
p(\mi|z) \propto 10^{-0.4(\mi-m_{i,\star}(z))(\alpha+1)} \exp\big(-10^{-0.4(\mi-m_{i,\star}(z))}\big), 
\label{eq:pmz}
\ee
where $\alpha$ is the faint-end slope of the Schechter luminosity function and $m_{i,\star}$ is the characteristic $i$-band magnitude of the red-sequence galaxies. Following \citet{redmap_des} and \citet{rozo2016}, we fix the parameter $\alpha=1$, and we calculate $m_{i, \star}(z)$ using the \textsc{EZgal}\footnote{\url{http://www.baryons.org/ezgal/}} (\citealt{ezgal_software,ezgal_paper}) implementation of the \citet{bc03} stellar population synthesis model. In the calculation of $m_{i,\star}(z)$ we also assume a solar metalicity, a Salpeter initial mass function (\citealt{chabrier2003}), and a single star formation burst at $z = 3$. Note that the argument of the exponential in Eq.~\ref{eq:pmz}  can be expressed in terms of luminosity ratios 
\be 
\frac{L}{L_{\star}} = 10^{-0.4(m_i-m_{i,\star}(z))}.
\label{eq:lratio}
\ee

Finally, the redshift prior takes the form of the derivative of the comoving volume with respect to redshift. This prior imposes uniformity of the comoving density across different redshifts.  
\begin{eqnarray}
p(z) & \propto & \frac{\mathrm{d}V_{\rm com}}{\mathrm{d}z} \label{eq:pz}\\
\frac{\mathrm{d}V_{\rm com}}{\mathrm{d}z} &=& (1+z)^{2}D_{A}^{2}(z)cH^{-1}(z),
\end{eqnarray} 
where $H(z)$ and $D_A(z)$ are the Hubble parameter and the angular diameter distance as a function of redshift $z$, respectively. 

The redshift prior takes into account the fact that for a given galaxy, the available volume is larger at higher redshifts. Therefore it ensures that the prior probability of finding a galaxy in a given redshift slice is proportional to the volume of that redshift slice. As a result, this choice of prior promotes a constant comoving density of galaxies across different redshifts. 

\subsection{Seed galaxies to estimate the red-sequence template}
\label{sec:seed}
Constructing the red-sequence template requires estimating the red-sequence ridge-line parameters as a function of redshift. Thus the first step is to find a set of seed red-sequence galaxies with secure spectroscopic redshifts to train the colour magnitude relation. 
In this work, we make use of the overlap between KiDS DR3 and the spectroscopic data from the thirteenth data release of Sloan Digital Sky Survey (hereafter SDSS DR13,~\citealt{sdss_dr13}) as well as the final spectroscopic data from the Galaxy And Mass Assembly survey (GAMA,~\citealt{driver2011}). These two datasets will be sufficient for selecting a set of seed galaxies needed for estimating the colour-magnitude relation. 

Creating the set of seed red galaxies is done by multiple filtering steps in the multi-dimensional colour-magnitude space, and in thin slices of redshift spanning the range $0.1<z<0.7$. 
The redshift range is limited by the available spectroscopic LRGs for training the red-sequence template as well as the wavelength range covered by the KiDS photometry. The $i$-band magnitude $m_i$ and three colour components $\{u-g,g-r,r-i\}$ used in our analysis are derived from  KiDS DR3 photometry and the spectroscopic redshifts $z_{\rm spec}$ are from GAMA and SDSS (see~\cref{sec:data}). 

First, we divide the dataset into thin redshift slices of $\Delta z = 0.02$ \footnote{We also experimented with other widths of the redshift slices ($\Delta z = 0.01$ , $\Delta z = 0.015$), and found no significant impact on the selection of seed galaxies for estimating the red-sequence ridge-line parameters.}. At each redshift slice, we fit two mixtures of Gaussian to the distribution of data points in the two dimensional (2D) space of $\{g-r,m_i\}$. One of the components of the Gaussian mixture model corresponds to the red population and the other component corresponds to the blue population\footnote{We have also repeated this step with a combination of $\{r-i,m_i\}$. We have noted that the choosing $r-i$ as the colour component in this step has no significant impact on the selection of seed galaxies.}. In particular, we employ the Extreme Deconvolution technique (hereafter XD, see \citealt{xd_code,xd_paper}) that finds the maximum likelihood estimates of the parameters of the mixture model in the cases where each data point has its own observed covariance matrix. That is, the XD model finds the underlying noise-deconvolved distribution of the heterogeneous dataset. In particular we make use of the $\textsc{astroml}$\footnote{\url{http://www.astroml.org}} implementation of XD (\citealt{astroml}).

In each slice of redshift, the data points are two dimensional vectors $\mathbf{x}_{\rm obs} = \{m_i,g-r\}$ 
and can be written as:

\be 
\mathbf{x}_{\rm obs} = \mathbf{x}_{\rm mod} + \mathrm{noise},
\label{eq:xdmod}
\ee
where $\mathbf{x}_{\rm mod}$ is the model described by the mixture of Gaussians, and the noise term is assumed to have a Gaussian distribution with zero-mean and a known covariance matrix $\mathbf{S}$:
\begin{eqnarray}
\mathbf{S} = 
\begin{bmatrix}
        \sigma_{i}^2      &                0   \\
 0        &         \sigma_{g}^2  +   \sigma_{r}^2  \\
 \end{bmatrix},
 \label{eq:Sgri}
\end{eqnarray} 
where $\sigma_g , \sigma_r, \sigma_i$ are photometric errors derived from KiDS DR3. The model vector 
$\mathbf{x}_{\rm mod}$ is drawn from a mixture of Gaussians with two components:
\be
p(\mathbf{x}_{\rm mod}) = \sum_{k=1}^{2} \pi_{k} \mathcal{N} \big(\mathbf{x}_{\rm mod} \; ; \; \boldsymbol{\mu}_{k}, \mathbf{V}_k \big),
\ee
where $\pi_k$, $\boldsymbol{\mu}_k$, and $\mathbf{V}_k$ are, respectively, the weight, the 2D mean vector, and the 2$\times$2 covariance matrix associated with the $k$-th Gaussian component, and 
\begin{eqnarray} 
\mathcal{N} \big(\mathbf{x}_{\rm mod} \; ; \; \boldsymbol{\mu}_{k}, \mathbf{V}_k \big) &=& \frac{\exp\Big(-\frac{1}{2}\Delta\mathbf{x}^{T}\mathbf{V}_k^{-1}\Delta \mathbf{x}\Big)}{\sqrt{(2\pi)^2\mathrm{det}(\mathbf{V}_{k})}}, \\
\Delta \mathbf{x} &=&\mathbf{x}_{\rm mod}-\boldsymbol{\mu}_{k}.
\end{eqnarray}
The component with larger mean $g-r$ corresponds to the red population. Then we select the points that are best represented by the 2D Gaussian distribution corresponding to the red population. 

Let us denote the mean and the covariance of the Gaussian component associated with red galaxies by $\boldsymbol{\mu}_r$ and $\mathbf{V}_r$, respectively. The first and the second components of $\boldsymbol{\mu}_r$ correspond to $m_i$ and $g-r$. Note that an initial estimate of the red-sequence ridge-line in the $\{m_i, g-r\}$ space can be found from $\boldsymbol{\mu}_r$ and $\mathbf{V}_r$:

\be 
(g-r)_{\rm mod} = \boldsymbol{\mu}_{r,2} + \mathbf{V}_{r,1,2}\big((m_i)_{\rm mod} - \boldsymbol{\mu}_{r,1}\big)/\mathbf{V}_{r,1,1},
\label{eq:init_line}
\ee
where $\boldsymbol{\mu}_{r,i}$ and $\mathbf{V}_{r,i,j}$ denote the $i$-th component of $\boldsymbol{\mu}_{r,i}$ and the $i,j$-th component of $\mathbf{V}_{r}$ respectively. Furthermore, the scatter $\sigma^{2}_{\rm mod}$ around this line can be defined in the following way:
\be 
\sigma^{2}_{\rm mod} = \mathbf{V}_{r,2,2} - \mathbf{V}^{2}_{r,1,2}/\mathbf{V}_{r,1,1}.
\label{eq:init_scatter}
\ee 
Combining Eqs.(\ref{eq:init_line},\ref{eq:init_scatter}) allows us to select data points in the $\{m_i,g-r\}$ space that are one sigma away from the initial estimate of the ridge-line. In other words, we keep the points that satisfy the following criteria
\be 
\big((g-r)_{\rm obs} - (g-r)_{\rm mod} \big)^{2} / (\sigma^{2}_{\rm mod} + \mathbf{S}_{2,2}) < 2,
\label{eq:chitwo}
\ee 
where $(g-r)_{\rm obs}$ is the observed colour, and $\mathbf{S}_{2,2}$, $(g-r)_{\rm mod}$, and $\sigma^{2}_{\rm mod}$ are given by Eqs.~(\ref{eq:Sgri},\ref{eq:init_line},\ref{eq:init_scatter}) respectively. Galaxies that meet this criteria~(\ref{eq:chitwo}) form an initial set of seeds for estimating the parameters of the red-sequence template. 

Furthermore, we employ a second filtering step. This is done in the 3D colour space $\{u-g,g-r,r-i\}$. Within narrow redshift intervals, red-sequence galaxies are expected to cluster in a compact volume of the colour space. 
If there exists a set of outlier galaxies that do not belong to the red population, the outlier galaxies are not going to follow the compact distribution of the red-sequence galaxies in the 3D colour space.Therefore, we can remove them by fitting a mixture of Gaussians with two components to the distribution of the remaining galaxies in the 3D colour space. One of the Gaussian components will capture the red population and the other Gaussian component will capture the outlier population. 

In this case, the observed data are 3D vectors $\mathbf{y}_{\rm obs} = \{u-g,g-r,r-i\}$, with observed uncertainties with zero mean and a known covariance $\tilde{\mathbf{S}}$:
\begin{eqnarray}
\tilde{\mathbf{S}} = 
\begin{bmatrix}
        \sigma_{u}^2 + \sigma_{g}^2      &    -\sigma_{g}^2    &         0   \\
 -\sigma_{g}^2       &         \sigma_{g}^2  +   \sigma_{r}^2 & -\sigma_{r}^2  \\
 0 & -\sigma_{r}^2 & \sigma_{r}^2  +   \sigma_{i}^2 \\
 \end{bmatrix}.
 \label{eq:Sugri}
\end{eqnarray} 
Once again we fit an XD model with two components to the distribution of the data in the colour space:
\be
p(\mathbf{y}_{\rm mod}) = \sum_{k=1}^{2} \tilde{\pi}_{k} \mathcal{N} \big(\mathbf{y}_{\rm mod} \; ; \; \tilde{\boldsymbol{\mu}}_{k}, \tilde{\mathbf{V}}_k \big),
\ee
where $\tilde{\pi}_k$, $\tilde{\boldsymbol{\mu}}_k$, and $\tilde{\mathbf{V}}_k$ are respectively the weight, the 3D mean vector, and the 3$\times$3 covariance matrix associated with the $k$-th Gaussian component: 
\begin{eqnarray} 
\mathcal{N} \big(\mathbf{y}_{\rm mod} \; ; \; \tilde{\boldsymbol{\mu}}_{k}, \tilde{\mathbf{V}}_k \big) &=& \frac{\exp\Big(-\frac{1}{2}\Delta\mathbf{y}^{T}\tilde{\mathbf{V}}_k^{-1}\Delta \mathbf{y}\Big)}{\sqrt{(2\pi)^3\mathrm{det}(\tilde{\mathbf{V}}_{k})}}, \\
\Delta \mathbf{y} &=&\mathbf{y}_{\rm mod}-\tilde{\boldsymbol{\mu}}_{k}.
\end{eqnarray}

Afterwards, we apply a cut based on the inferred mean vectors of the Gaussian distributions. The mean of the Gaussian component capturing the red (outlier) galaxy population has a higher (lower) mean along the $r-i$ axis. We denote the mean and the covariance of the Gaussian component with a higher mean along the $r-i$ axis with $\tilde{\boldsymbol{\mu}}_{r}$ and $\tilde{\mathbf{V}}_{r}$ respectively. Finally, we select those galaxies that, in the 3D colour space, are within one sigma from the mean of the Gaussian component corresponding to the red population. That is, galaxies must meet the following criteria in order to be considered in the collection of seed galaxies for training the template model:
\be 
\big(\mathbf{y}_{\rm obs} - \tilde{\boldsymbol{\mu}}_{r} \big)^{T}\big(\tilde{\mathbf{S}}+\tilde{\mathbf{V}}_r\big)^{-1}\big(\mathbf{y}_{\rm obs} - \tilde{\boldsymbol{\mu}}_{r}\big) < 2.
\label{eq:chithree}
\ee

The conditions~(\ref{eq:chitwo},\ref{eq:chithree}) ensure that only the galaxies in the core of the red-sequence population of galaxies are considered as seeds for inferring the colour magnitude relation.

\subsection{Red-sequence template}

Now we discuss how we estimate the parameters of the red-sequence template~(\ref{eq:temp}) with the seed galaxies. The template is fully specified by the parameters $\va,\vb,\vCi$, as well as by the reference $i$-band magnitude $m_{i,\rm ref}(z)$. 

%The choice of $m_{i,\rm ref}(z)$ depends on the investigator. 
We choose to estimate the parameter $m_{i,\rm ref}(z)$ from $\mathtt{Cubic Spline}$ interpolation of a set of $m_{i,\rm ref}$ parameters at some Spline nodes uniformly distributed between $z=0.1$ and $z=0.7$. The Spline nodes are chosen to be the midpoints in the redshift intervals that were used to select the seed red galaxies. We also select $\boldsymbol{\mu}_{r,1}$ as our choice of $m_{i,\rm ref}$ at the Spline nodes.

Moreover, we also choose to parametrize $\va$, $\vb$, $\vCi$ by specifying discrete Spline nodes at different redshifts. We note that the only parameter that varies significantly in short redshift intervals is $\va$. 
Thus for $\va$ we choose Spline nodes with spacings of $\Delta z = 0.05$ uniformly distributed between $z=0.1$ and $z=0.7$. For $\vb$ and $\vCi$ however, wider spacings for the Spline nodes are chosen (see~\citealt{redmap_sdss}). In our work, spacing of $\Delta z = 0.1$ and $\Delta z = 0.14$ are chosen for the Spline nodes at which we parametrize $\vb$ and $\vCi$. 

Furthermore, as discussed earlier, we decide to ignore the off-diagonal elements of the intrinsic covariance matrix. 
Therefore, there are three parameters at every intrinsic invariance Spline node, three parameters at every slope Spline node, and three parameters at every intercept Spline node. We denote the multi-dimensional vector representing these parameters as $\bm{\theta}$. The vector $\bm{\theta}$ can be estimated by minimizing the objective function:
\be 
\mathcal{O}(\bm{\theta}) = -2\sum_{j=1}^{N_{\rm gal}} \ln \; p(\vc_j|m_{i,j},z_j;\bm{\theta}), 
\label{eq:otheta}
\ee
where the summation is over all seed galaxies and the conditional probability $p(\vc_j|m_{i,j},z_j;\bm{\theta})$ for $j$-th galaxy is evaluated using Eq.~\ref{eq:temp}. Minimization of the objective function (\ref{eq:otheta}) is done by the $\mathtt{scipy}$ implementation of the $\mathtt{BFGS}$ algorithm \citep{bfgs}.

\subsection{Initial redshift estimation}

Given the red-sequence template (Eq~\ref{eq:temp}), the magnitude distributions (Eq~\ref{eq:pmz}), and redshift priors (Eq~\ref{eq:pz}),
one can optimize $p(z|m,\vc)$ to obtain a maximum a posteriori estimate $\hat{z}$ of the red-sequence redshift of galaxies. In practice, we use the $\mathtt{scipy}$ implementation of the $\mathtt{BFGS}$ optimizer to minimize the following objective function:

\begin{eqnarray} 
-2\ln \; p(z|m_i,\vc) &=& \chi^{2}_{\rm red}(z) + \ln\; \mathrm{det}\big(\vCt\big) \nonumber \\  
                    &-& 2 \ln \; \Big|\frac{\mathrm{d}V}{\mathrm{d}z}\Big|-2\ln\;p(m_i|z).
\end{eqnarray}
Therefore, an estimate of redshift $\hat{z}$ can be found according to
\be 
\hat{z} = \mathrm{argmin}_{z} \; \big[-2\ln \; p(z|m_i,\vc)\big],
\label{eq:minim}
\ee
where $\mathrm{argmin}_{z} \; \big[-2\ln \; p(z|m_i,\vc)\big]$ is the value of $z$ that minimizes the function: $-2\ln \; p(z|m_i,\vc)$.
\subsection{Selection criteria}
Once we have an estimate of the redshifts of LRG candidates, we can apply appropriate 
cuts to the catalogue to obtain a sample of luminous red-sequence galaxies. LRG candidates need 
to meet two criteria in order to pass the cuts. First, we apply a cut based on the maximum red-sequence chi-squared $\chi^{2}_{\rm red}({\hat{z}})$ achieved by minimizing the objective function~(\ref{eq:minim}). That is, at a given redshift, if $\chi^{2}_{\rm red}({\hat{z}})$ is less than a specified maximum allowable chi-squared $\chi^{2}_{\rm max}(z)$, the LRG candidate passes the chi-squared  criterion. We will postpone discussion of estimating $\chi^{2}_{\rm max}(z)$ to Section~\ref{sec:chimax}.

The chi-squared criterion ensures that the selected galaxies belong to the red-sequence population. In other words, it ensures that the selected galaxy colours and magnitudes are well-described by the inferred red-sequence template.
As we are mainly interested in the luminous red galaxies, we impose another cut that selects galaxies that are more luminous than a certain threshold. In section \ref{sec:overview}, we defined the luminosity ratio $l = L/L_{\star}$(see Eq.~\ref{eq:lratio}). At a given redshift, we only select galaxies with $l>l_{\rm min}$, or equivalently with $L > L_{\rm min} = l_{\rm min }L_{\star}$. 

As we discuss later in section~\ref{sec:chimax}, we will construct two samples: a high density sample with $l_{\rm min}=0.5$ and a luminous sample with $l_{\rm min}=1$. 

\subsection{Photo-z afterburner}\label{sec:afterburner}

A set of LRG candidates with secure spectroscopic redshifts can be used to calibrate the photometric redshifts obtained by our method. In practice, we only make use of a subset of LRG candidates with spectroscopic redshifts and we leave the rest for validation. The calibration set consists of randomly selected 50\% of the galaxies in the overlap of KiDS DR3 with SDSS DR13, GAMA, and 2dFLenS (see section~\ref{sec:data} for data details).   

We assume that the calibration can be parametrized by a redshift offset parameter $\delta z$ that is a smooth function of redshift, $\delta z = \delta z(\hat{z})$. In order to estimate $\delta z(\hat{z})$
we choose a set of ten Spline nodes $\{z_i\}_{i=1}^{10}$ uniformly spaced between $z=0.1$ and $z=0.7$. Then the task of estimating $\delta z$ is reduced to the task of estimating $\delta z (z_i)$ for $i=1,...,10$, where $\delta z (z_i)$ is $\delta z$ evaluated at the spline node $z_i$.

In order to estimate $\delta z(z_i)$, we construct the following objective function:

\be 
E\big(\{\delta z_i\}\big) = \sum_{\tilde{z}_{\rm spec}} |\tilde{z}_{\rm spec} - \delta z(\hat{z}) - \hat{z}|,
\label{eq:lcalib}
\ee 
where the summation is over spectroscopic redshifts of galaxies in the calibration sample. 

Note that in Eq.~\ref{eq:lcalib} we have used an $L_1$ norm\footnote{For a given vector $y$, consisting of target values of a given quantity, and a vector $\hat{y}$, composed of the estimates of the same quantity, the $L_1$ cost function is defined as the sum over the absolute values of the differences: $L_1(y,\hat{y}) = \sum_{i}|y_i - \hat{y}_i|$. Similarly, an $L_2$ cost function is given by the sum over the squared-differences: $L_2(y,\hat{y}) = \sum_{i}(y_i - \hat{y}_i)^2$.}  for the objective function $E$. The motivation for our choice of $L_1$ norm is that it is more robust against outliers. If there is a fraction of galaxies with highly biased redshift estimates, they could bias our estimate of $\delta z(z)$. Using a conventional $L_2$ norm in the objective function $E$ can be more sensitive to these outliers. Therefore, in order to reduce the sensitivity of our redshift calibration method to outliers we use an $L_1$ norm instead. 

As we point out in section~\ref{sec:chimax}, this redshift calibration scheme is done within the $\chi_{\rm max}^{2}(z)$ calibration. This is due to the fact that both luminosity ratios $l(z)$ and the red-sequence chi-squared values $\chi^{2}_{\rm red}(z)$ of LRG candidates depend on the estimated red-sequence redshifts. 
After every redshift calibration ($\hat{z} \rightarrow \hat{z} + \delta z(\hat{z})$), the values of $l(\hat{z})$ and $\chi^{2}_{\rm red}(\hat{z})$ need to be updated as well. For this reason, the entire photo-z afterburner operation needs to be performed within calibration of maximum allowable chi-squared $\chi^{2}_{\rm max}(z)$ which we will now explain. 

%We also remind the readers that only a subset of LRG candidates with spectroscopic redshifts are used for this purpose. The rest of the LRG candidates are used for the validation of our method. 

\subsection{Calibration of red-sequence chi-squared}\label{sec:chimax}

We estimate the redshift-dependent $\chi^{2}_{\rm max}$ by requiring the final red-sequence sample to have nearly constant comoving density across cosmic time. In other words, we require the number of LRGs to be proportional to the comoving volume available for them.
This can be done by counting the number of LRG candidates in narrow bins of redshift and then comparing this number with the expected number assuming a constant comoving density. 

Let us denote the fraction of sky covered by the survey by $f_{\rm s}$. Then for a given comoving number density $\bar{n}$, the expected number of LRGs in a redshift interval $\Delta z_j$ centred on redshift $z_j$ is 
\be 
N_j \simeq \bar{n}f_{\rm s}\frac{\mathrm{d}V_c}{\mathrm{d}z}(z_j)\Delta z_j,
\label{eq:nj}
\ee
where $\frac{dV_c}{dz}(z_j)$ is the derivative of the comoving volume with respect to redshift evaluated at $z_j$. The number of LRG candidates in the redshift interval $\Delta z_j$ will be denoted as $H_j$. Given a specified minimum luminosity ratio $l_{\rm min}=L_{\rm min}/L_{\star}$, the number count $H_j$ depends on the number of galaxies that pass the requirement $\chi^{2}_{\rm red}(z_j) < \chi^{2}_{\rm max}(z_j)$. 

As a result, one needs to adjust the values of $\chi^{2}_{\rm max}(z_j)$ so that for a given choice of the luminosity ratio, $H_j$ matches the prediction based on constant comoving number density $N_j$~(Eq.~\ref{eq:nj}). We choose to model $\chi^{2}_{\rm max}$ as a smooth function of redshift. Thus, we choose to parametrize it by selecting a few Spline nodes $z_{k}$ uniformly spaced between $z=0.1$ and $z=0.7$, and then interpolating the values of $\chi^{2}_{\rm max}(z_k)$ to a given redshift $z_j$ using $\mathtt{CubicSpline}$ interpolation.

We estimate the set of parameters $\chi^{2}_{\rm max}(z_k)$ by minimizing the following objective function:

\be 
\mathcal{O}\big(\{\chi^{2}_{\rm max}(z_k)\}\big) = \sum_j \; \frac{(H_j-N_j)^{2}}{(H_j + N_j)},
\label{eq:hist}
\ee
where the denominator is simply given by the Poisson noise calculated from the galaxy number counts $H_j$ and the expected number counts assuming constant density $N_j$. Note that in evaluation of Eq.~\ref{eq:hist} we use a more fine binning than the Spline nodes at which we parametrize $\chi^{2}_{\rm max}$. 

In section~\ref{sec:afterburner} we discussed our strategy for estimating the calibration errors as a 
function of redshift. Estimating $\chi^{2}_{\rm max}(z)$ through iterative minimization of the objective function (Eq.~\ref{eq:hist}) is based on the assumption that the redshifts are calibrated since both $L/L_{\star}$ and $\chi^{2}_{\rm red}(z)$ are modified after calibration of redshifts. Therefore, before evaluating the objective function $\mathcal{O}\big(\{\chi^{2}_{\rm max}(z_k)\}\big)$ at each iteration, the afterburner procedure is performed, and the luminosity ratios $L/L_{\star}$ and the red-sequence chi-squared values $\chi^{2}_{\rm max}$ are updated for all the galaxies in the survey. Afterwards, given a choice of luminosity ratio and the $\chi^{2}_{\rm max}(z_k)$, the objective function (Eq.~\ref{eq:hist}) is evaluated. 

We run the initial redshift estimation and $\chi^{2}_{\rm red}$ calculation for all objects in the photometric catalogue. Prior to the calibration of the red-sequence chi-squared, we set an upper limit for the apparent $i$-band magnitude of the objects in the catalogue. For chi-squared calibration of objects with $l_{\rm min} = 0.5$ we set the maximum $m_i$ to 21.6, and for objects with $l_{\rm min} = 1$ we set the maximum $m_i$ to 20.8. 
These upper limits ensure that the red-sequence photo-$z$ scatters are under control (less than $\sim$ 0.016), while the final catalogue has the desired constant comoving density. Hereafter, the final calibrated red-sequence redshifts are denoted by $z_{\rm red}$.

%%%%%%%%%%%%%%%%%%%%%%%%%%%%%%%%%%%%%%%%%%%%%%%%%%%%%%%%
% Figure: SKY
%%%%%%%%%%%%%%%%%%%%%%%%%%%%%%%%%%%%%%%%%%%%%%%%%%%%%%%%
%\begin{figure*}
%\begin{center}
%\includegraphics[width=\textwidth]{figures/sky_north.png}
%\caption{\label{fig:sky_n} Distribution of the \texttt{dense} LRG sample with $L/L_{\star}>0.5$ in the patch of sky in the equatorial fields covered by KiDS DR3.}
%\end{center}
%\end{figure*}
%\begin{figure*}
%\begin{center}
%\includegraphics[width=\textwidth]{figures/sky_south.png}
%\caption{\label{fig:sky} Distribution of the LRGs selected photometrically from KiDS DR3 for the \texttt{dense} sample with $L/L_{\star}>0.5$ in the Equatorial fields (Top panel) and the Southern Galactic cap (Bottom Panel)} 
%\end{center}
%\end{figure*}

\section{Photometric redshifts}\label{sec:photoz}
\subsection{Selection summary}

As our final selection step, we decide to construct two samples, with minimum $L/L_{\star}$ 
ratios of 0.5 and 1. Furthermore, in the $\chi^{2}_{\rm max}(z)$ calibration, 
we choose to keep the comoving density of each sample fixed. We call these two samples the $\mathtt{dense}$ sample and the $\mathtt{luminous}$ sample. The $\mathtt{dense}$ sample has a mean comoving density of $10^{-3}\;h^{3}\mathrm{Mpc}^{-3}$ and a minimum $L/L_{\star}$ of 0.5. On the other hand, the $\mathtt{luminous}$ sample has a mean comoving density of $2\times10^{-4}\;h^{3}\mathrm{Mpc}^{-3}$ and a minimum $L/L_{\star}$ of 1. The selection is summarized in Table~\ref{tab:prior}.

%The estimated maximum red-sequence chi-squared for the two samples are shown in Figure \ref{fig:chi}.  

%%%%%%%%%%%%%%%%%%TABLES%%%%%%%%%%%%

\begin{table}
	\centering
	\caption{{\bf LRG sample selection summary}: The LRG sample selected from KiDS DR3 and the corresponding luminosity thresholds and comoving number densities. The density parameters are in unit of $h^{3}\;\mathrm{Mpc}^{-3}$. The redshift range of both samples is $z_{\rm red} \in [0.1,0.7]$.}
	\label{tab:prior}
	\begin{tabular}{lcccr} % four columns, alignment for each
		\hline
		LRG Sample & $L_{\rm min}/L_{\star}$ & number density & total number\\
		\hline
		$\mathtt{dense}$ & 0.5 & $10^{-3}$ & $191,775$\\
		$\mathtt{luminous}$ & 1 & $2\times 10^{-4}$ & $38,671$\\
		\hline
	\end{tabular}
\end{table}

\begin{figure*}
 \begin{tabular}{cc}
\includegraphics[width=\columnwidth]{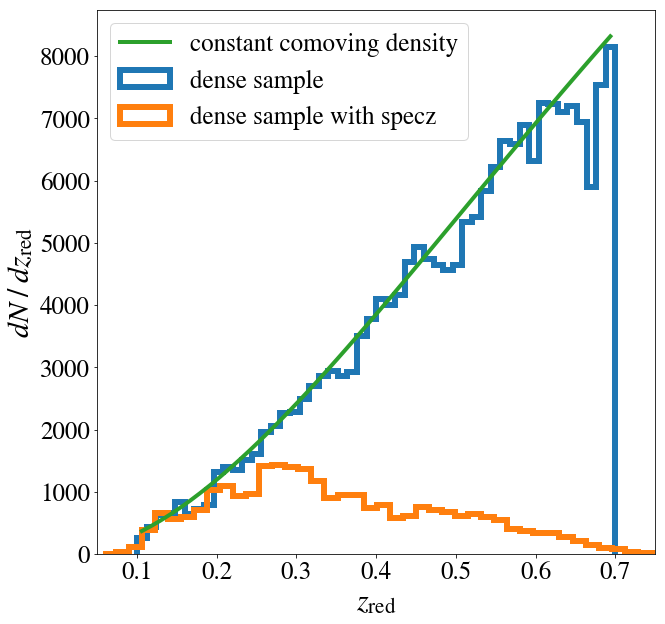}
\includegraphics[width=\columnwidth]{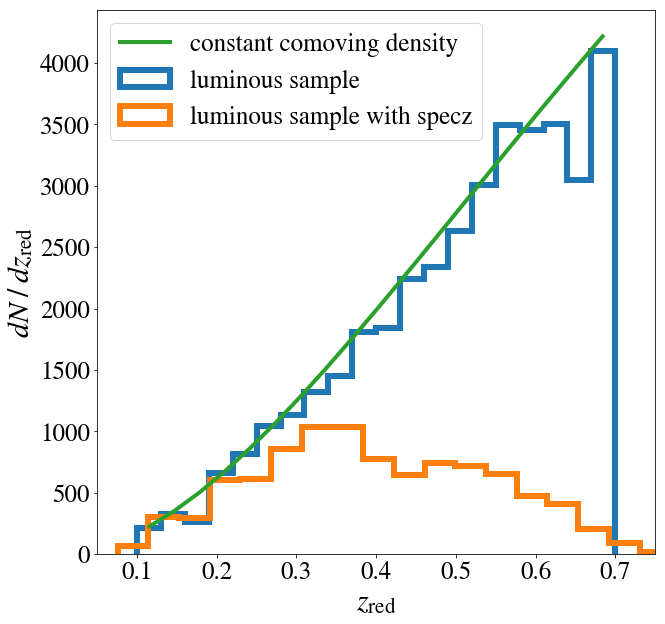}
\end{tabular}
\caption{\label{fig:dndz} Histogram of the redshift distribution of the photometrically selected luminous red galaxies in KiDS based on the method described in this paper. Left: Comparison between the distribution of galaxies in the $\texttt{dense}$ sample (blue histogram) and the galaxies in the $\texttt{dense}$ sample with secure spectroscopic redshifts (orange histogram). The green curve shows the expected distribution assuming a constant comoving density of $n = 10^{-3} \; h^{3}\;\mathrm{Mpc}^{-3}$. Right: same as the left panel but for the galaxies in the $\texttt{luminous}$ sample, with the green line showing the expected redshift distribution assuming constant a comoving density of $n = 2 \times 10^{-4} \; h^{3}\;\mathrm{Mpc}^{-3}$. There is a good agreement between the redshift distribution of the selected galaxies in both samples and the expected distributions based on the assumption of constant comoving density.}
\end{figure*}

Figure~\ref{fig:dndz} shows the comparison between the redshift distribution of our selected red galaxies (solid blue histogram) and the expected distribution based on the assumption of constant comoving density (solid green line). 
The left panel of Fig.~\ref{fig:dndz} shows the redshift distribution of galaxies in the $\dense$ sample while the right panel shows that of the galaxies in the $\lum$ sample. Also shown in Fig.~\ref{fig:dndz} are the redshift distributions of the selected red galaxies with spectroscopic redshifts (solid orange histograms). 

\begin{figure*}
 \begin{tabular}{cc}
\includegraphics[width=0.33\textwidth]{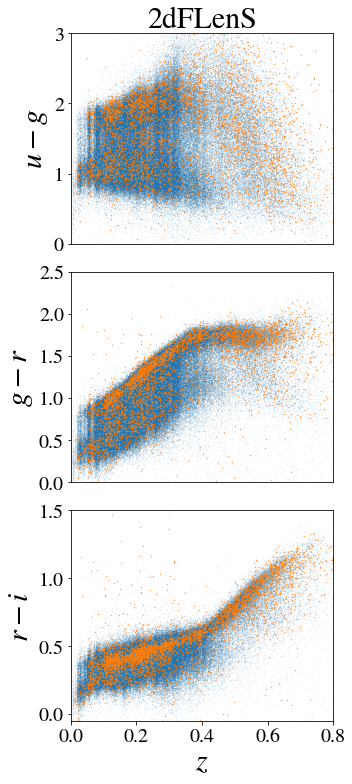}
\includegraphics[width=0.33\textwidth]{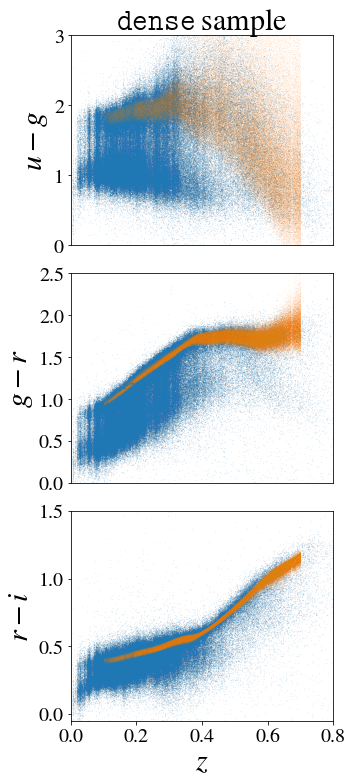}
\includegraphics[width=0.33\textwidth]{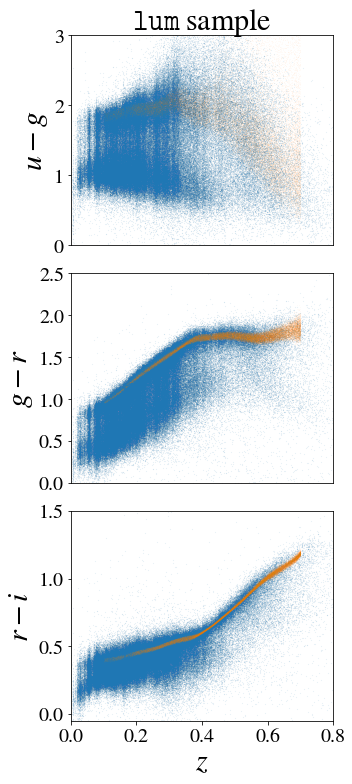}
\end{tabular}
\caption{\label{fig:2df} Left column: The redshift dependence of the colours of SDSS+ GAMA galaxies with spectroscopic redshifts (blue points) used in this study and that of the 2dFLenS galaxies (orange points). Shown from Top to Bottom are the redshift dependence of $u-g$, $g-r$, and $r-i$. Middle column: Same as the Left column with the exception that the over-plotted orange points are the galaxies in the $\mathtt{dense}$ sample, and the redshifts are the estimated red-sequence redshifts of these galaxies. Right column: same as the Middle column but with the orange points showing the galaxies in the $\mathtt{luminous}$ sample. For better visibility the points corresponding to the 2dFLenS galaxies are chosen to be much larger than the points corresponding to galaxies in the $\mathtt{dense}$ and the $\mathtt{lum}$ samples.}
\end{figure*}

We note that in general there is a good agreement between the redshift distribution of the selected red galaxies and the expected distribution based on constant comoving density. The number of selected LRGs at higher redshifts is significantly higher than that of LRGs with secure spectroscopy. This demonstrates how the method presented in this work can exploit the information available in the red-sequence template in order to select a well-controlled sample of galaxies in a wide range of redshifts. Figure~\ref{fig:2df} shows the distribution of colours versus redshift for SDSS and GAMA galaxies (blue points) versus the distribution of galaxies in the 2dFLenS luminous red galaxy survey (left column, orange points), galaxies in the $\dense$ sample (middle column, orange points), and galaxies in the $\lum$ sample (middle column, orange points). We note that compared to the 2dFLenS galaxies, the galaxies selected in this work sample the red-sequence more continuously.

\begin{figure*}
 \begin{tabular}{cc}
\includegraphics[width=\columnwidth]{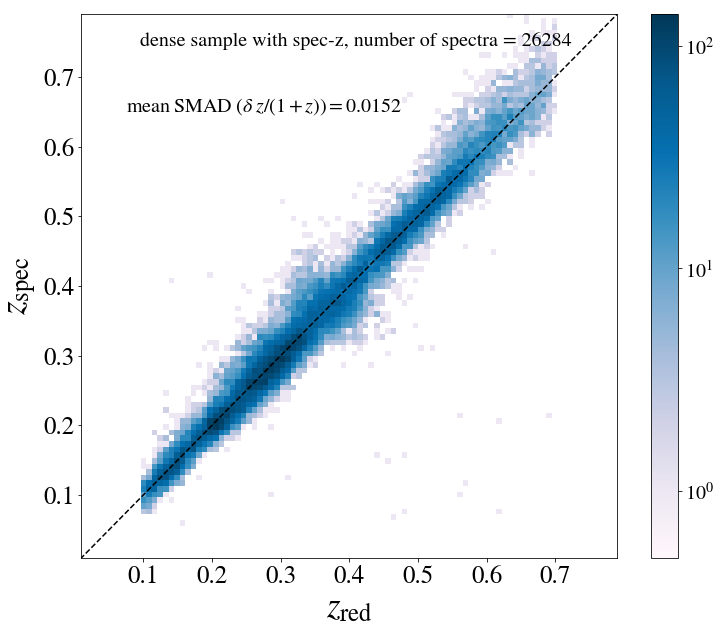}
\includegraphics[width=\columnwidth]{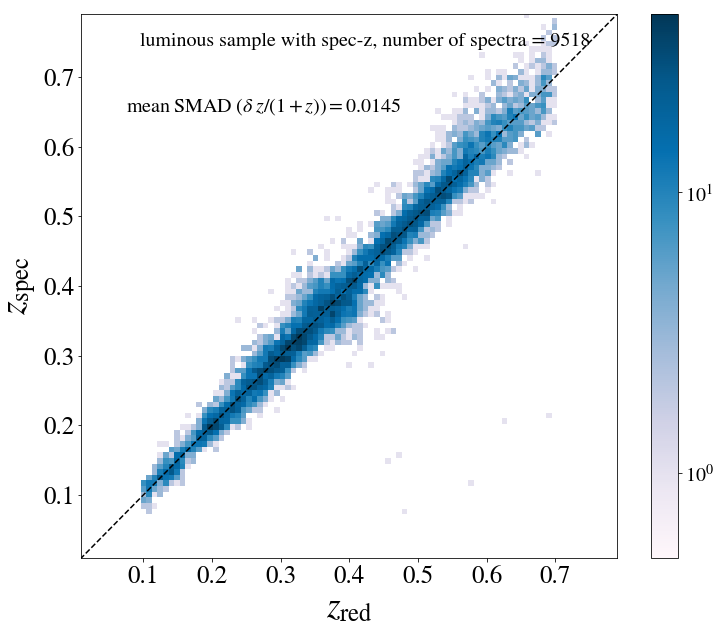}
\end{tabular}
\caption{\label{fig:sigma} Left panel: Demonstration of the performance of the estimated red-sequence redshifts $z_{\rm red}$ of galaxies in the $\dense$ sample. 
The heat map demonstrates the red-sequence redshifts ($x$-axis) versus the spectroscopic redshifts ($y$-axis). Right panel: Same as the left panel but showing the red-sequence redshift performance of galaxies in the $\lum$ sample. In both panels the dashed line shows the $z_{\rm spec} = z_{\rm red}$ line.}
\end{figure*}

\begin{figure}
%\begin{tabular}{cc}
%\includegraphics[width=\columnwidth]{figures/bias_dense.png}
%\includegraphics[width=\columnwidth]{figures/bias_lum.png}
%\end{tabular}
\includegraphics[width=\columnwidth]
{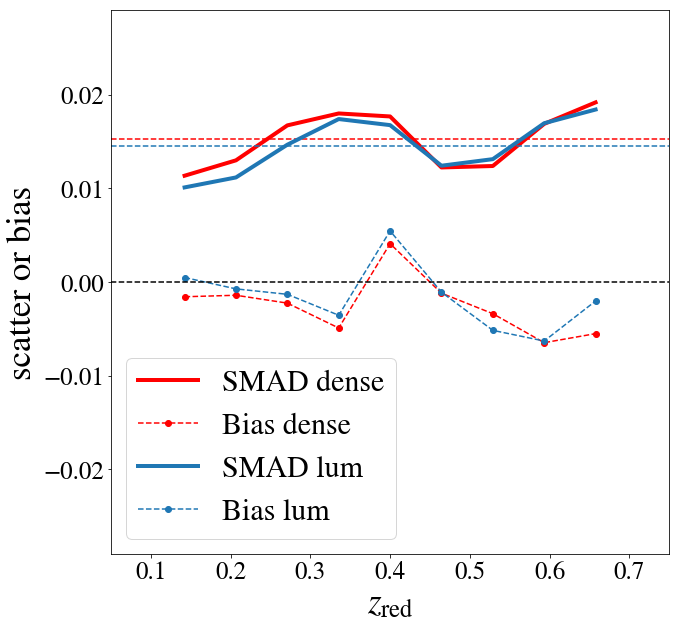}
\caption{\label{fig:bias_scatter} Bias and scatter of the estimated red-sequence redshifts of galaxies in the $\mathtt{dense}$ sample (Red) and in the $\mathtt{luminous}$ sample (Blue) as a function of $z_{\rm red}$. The scatter (solid line) is the standard median absolute deviation (SMAD) of the quantity $(z_{\rm red}-z_{\rm spec})/(1+z_{\rm red})$ measured in bins of redshift. The bias (dashed dotted line) is given by the mean of $\delta z = z_{\rm red} - z_{\rm spec}$ measured in bins of redshift. The dashed lines show the mean of the estimated binned scatters. We note that the scatter is nearly constant as a function of redshift and its mean value is approximately 0.015 (0.014) for the $\dense$ ($\lum$) sample. Moreover, the bias is always smaller than the predicted scatter.}
\end{figure}

\begin{figure}
% \begin{tabular}{cc}
\includegraphics[width=\columnwidth]{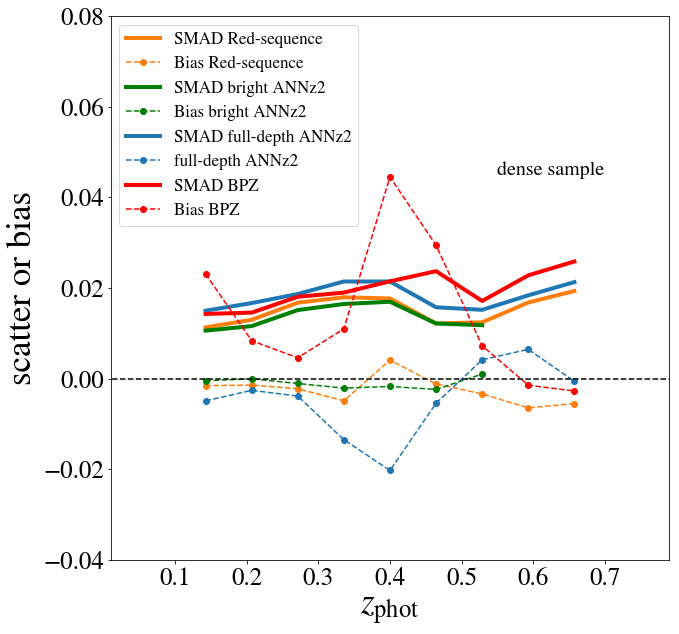}
%\includegraphics[width=\columnwidth]{figures/bias_lum_all.png}
%\end{tabular}
\caption{\label{fig:comparison} Comparison between the performances of red-sequence photo-$z$'s (shown in orange), bright ANNz2 photo-$z$'s (shown in green), full-depth ANNz2 photo-$z$'s (shown in blue), and BPZ photo-$z$'s (shown in red) for galaxies in the $\mathtt{dense}$ LRG sample. Scatter is estimated by calculating the SMAD of $(z_{\rm phot}-z_{\rm spec})/(1+z_{\rm spec})$ and is shown by solid lines, while bias $\delta z = z_{\rm phot} - z_{\rm spec}$ is shown with points.
Both bias and scatter are calculated in bins of redshifts. We note that the estimated scatters from the all methods are very similar with the bright ANNz2 and red-sequence photo-$z$'s having the best performances.}
\end{figure}

\begin{table*}
	\centering
	\caption{{\bf Photo-z performance comparison:} 
    Comparison between the performances of four photo-$z$ estimation methods when applied to galaxies in the $\dense$ sample. The quantities considered here are the bias defined as $|z_{\rm phot} - z_{\rm spec}|$, scatter defined as 1.4826 times the median-absolute-deviation of $(z_{\rm phot} - z_{\rm spec})/(1+z_{\rm spec})$, the percentage of 5$\sigma$ outlier fraction, and the percentage of catastrophic outliers. We define the percentage of catastrophic outliers as the percentage of galaxies for which $|z_{\rm phot} - z_{\rm spec}|/(1+z_{\rm spec})>0.15$. The first three quantities are computed in bins of redshift and then the means of the binned values are reported in the Table.}
	\label{tab:comparison}
	\begin{tabularx}{0.85\textwidth}{lccccr} % four columns, alignment for each
		\hline
		Photo-$z$ estimation method & |Bias| & Scatter & 5$\sigma$ outlier fraction (\%) & Catastrophic outlier fraction (\%)\\
		\hline
		$\mathtt{Red}$-$\mathtt{sequence}$ & $3.4\times 10^{-3}$ & $0.0152 $ & 1.3&0.05\\
		$\mathtt{Bright \; ANNz2}$ & $1.3\times 10^{-3}$ & $0.0135$&0.9 & 0.04\\
        $\mathtt{Full}$-$\mathtt{depth \; ANNz2}$ & $6.8\times 10^{-3}$ & $0.0182$&1.5 & 0.18\\
        $\mathtt{BPZ}$ &$14.7\times 10^{-3}$ & 0.0196 &4.2 & 0.06\\
		\hline
	\end{tabularx}
\end{table*}

%%%%%%%%%%%%%%%%%%%%%%%% PHOTOZ SYSTEMATICS %%%%%%%%%%%%%%%%%%%%%%%%

\begin{figure*}
 \begin{tabular}{cc}
\includegraphics[width=0.33\textwidth]{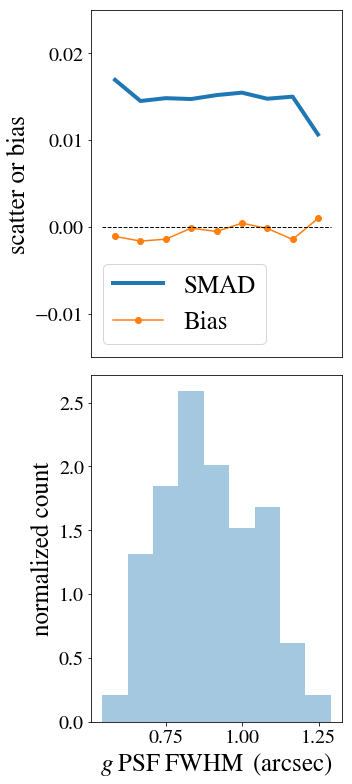}
\includegraphics[width=0.33\textwidth]{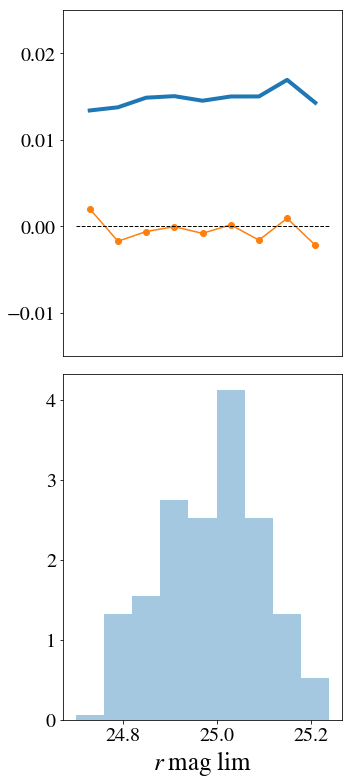}
\includegraphics[width=0.33\textwidth]{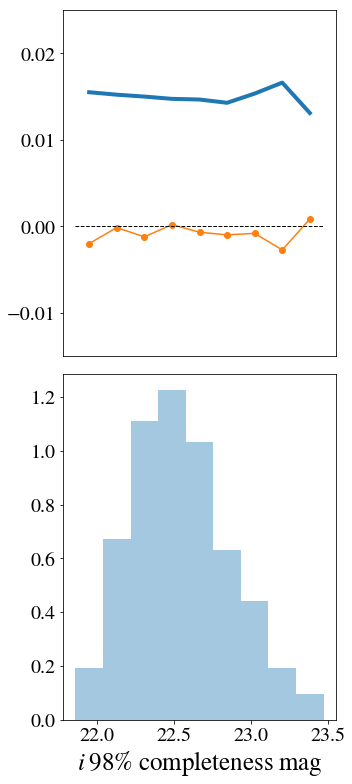}
\end{tabular}
\caption{\label{fig:dense_psf} Dependence of the red-sequence photo-$z$ errors of galaxies in the $\mathtt{dense}$ sample on the survey systematics. Left panel: photo-$z$ error as a function of PSF FWHM (in units of arcseconds) in the $g$ band. The seeing values are the mean PSF FWHM of the coadded images of the survey tiles in KiDS-DR3. Middle panel: photo-$z$ error as a function of the the limiting magnitude (2$\sigma$ in 2 arcsecond aperture) in the $r$ band. The limiting magnitudes are the mean values calculated from the coadded images of the survey tiles in KiDS-DR3. Right Panel: photo-$z$ error as a function of the 98\% completeness magnitudes in the $i$ band. These values are obtained from the single band source list in KiDS DR3 and represent the mean completeness magnitudes of the survey tiles.  Both red-sequence photo-$z$ scatter and bias are nearly constant functions of the survey systematics.}
\end{figure*}

%\begin{figure*}
%\includegraphics[width=\textwidth]{new_figures/psf_dense.png}
%\caption{\label{fig:dense_psf} Dependence of the photo-$z$ errors of galaxies in the $\mathtt{dense}$ sample on the PSF FWHM (in units of arcseconds) in the $gri$ bands. The seeing values are the mean PSF FWHM of the coadded images of the survey tiles in KiDS-DR3.  We note that both photo-$z$ scatter and bias are nearly constant functions of the mean PSF FWHM of the survey tiles.}
%\end{figure*}

%\begin{figure*}
%\includegraphics[width=\textwidth]{new_figures/comp_dense.png}
%\caption{\label{fig:dense_comp} Dependence of the photo-$z$ errors of galaxies in the $\mathtt{dense}$ sample on the 98\% completeness magnitudes in the $gri$ band. These values are obtained from the single band source list in KiDS DR3 and represent the mean completeness magnitudes of the survey tiles.  We note that both photo-$z$ scatter and bias are nearly constant functions of the mean 98\% completeness magnitudes of the survey tiles.}
%\end{figure*}

%\begin{figure*}
%\includegraphics[width=\textwidth]{new_figures/maglim_dense.png}
%\caption{\label{fig:dense_maglim} Dependence of the photo-$z$ errors of galaxies in the $\mathtt{dense}$ sample on the limiting magnitude (2$\sigma$ in 2 arcsecond aperture) in the $gri$ bands. The limiting magnitudes are the mean values calculated from the coadded images of the survey tiles in KiDS-DR3. We note that both photo-$z$ scatter and bias are nearly constant functions of the mean limiting magnitudes of the survey tiles.}
%\end{figure*}

\subsection{Redshift performance}\label{sec:performance}

We will now verify the performance of LRG red-sequence photo-$z$'s using the overlapping spectroscopy. As already mentioned in section~\ref{sec:spec}, the spec-$z$'s originate from SDSS DR13, GAMA, and 2dFLenS.
Figure~\ref{fig:sigma} shows the performance of the estimated red-sequence redshifts for the $\dense$ sample (left panel) and the $\lum$ sample (right panel). 
In general, there is an excellent agreement between the red-sequence redshifts and the spectroscopic redshifts. But in order to asses the quantitative performance of the estimated redshifts for the selected LRGs we make use of two quantities in bins of $z_{\rm red}$. The first quantity is the mean bias $\delta z = z_{\rm red} - z_{\rm spec}$ in bins of $z_{\rm red}$. The second quantity is the scatter which is estimated via the standard median absolute deviation (SMAD) of $(z_{\rm red} - z_{\rm spec})/(1+z_{\rm spec})$ in bins of $z_{\rm red}$. 

Figure~\ref{fig:bias_scatter} shows the mean bias and SMAD for galaxies in the $\mathtt{dense}$ sample (red) and those in the $\mathtt{luminous}$ sample (blue). All quantities are measured in bins of $z_{\rm red}$. The mean scatter of the red-sequence redshifts of $\lum$ and $\dense$ galaxies is 0.0145 and 0.0152 respectively, and the mean absolute value of bias is respectively $2.9\times 10^{-3}$ and $3.4\times 10^{-3}$. In general the estimated scatter is nearly constant but higher at the redshifts corresponding to the transition of the 4000 Angstrom break between the photometric filters. Note that the estimated bias is also higher at those redshifts. 

The estimated $z_{\rm red}$ scatters of the selected red-sequence galaxies is limited by using the broad-band KiDS photometry. The mean $z_{\rm red}$ scatters of the $\dense$ and the $\lum$ sample are very similar. That is due to the fact that a large fraction of $\dense$ galaxies with spectroscopy are luminous ($L/L_{\star}>1$). The 5-$\sigma$ outlier fraction of both samples are about 1\% and the catastrophic outlier fractions are about 0.1\%. After investigating the spectroscopy of the red-sequence galaxies, we have noted that at fixed photometric redshift bins, the outlier galaxies have slightly higher H$\alpha$ fluxes than the non-outlier galaxies. That may suggest that a residual star formation in the outlier galaxies make them appear bluer than the non-outlier galaxies.

\subsection{Comparison with other methods}
For the selected galaxies, we also assess the quality of the estimated red-sequence redshifts by comparing them with other photo-$z$ estimation methods available in KiDS DR3. These include template-fitting BPZ photo-$z$'s (\citealt{bpz1999}), %(\citealt{kids_dr3,hendrick2017}) 
as well as those determined by the machine learning method ANNz2 (\citealt{sadeh2016}), as described in \citet{kids_dr3} and \cite{kids_annz}. Those photo-$z$'s are available for all galaxy types, but here we will discuss their performance only for the LRGs contained in our samples. As in the previous section, we will also employ overlapping spectroscopy to derive photo-$z$ performance metrics.

For the machine-learning results, we make use of two estimates of ANNz2 photo-$z$'s presented in \citet{kids_annz}. The first set of redshifts, which we call the bright ANNz2 photo-$z$'s, are the photo-$z$'s that are exclusively trained on GAMA, and their performance is enhanced by using not only magnitudes, but also colours and angular sizes in the feature space. After a posteriori cut to the apparent magnitude ($m_{\rm r,AUTO}<20.3$), \citet{kids_annz} demonstrated that compared to the GAMA spectroscopic redshifts, the bright ANNz2 photo-$z$'s have a scatter of 0.026 and a mean bias of $-3.3 \times 10^{-3}$. Note that as a result of the magnitude cut, the maximum redshift in the bright ANNz2 catalogue is approximately 0.62. The second machine learning-based catalogue, which we call the full-depth ANNz2 catalogue, consists of photo-$z$'s that are trained on the full depth of KiDS DR3 data exploiting the overlapping deep spectroscopic samples. For those photo-$z$'s only GAaP magnitudes were used as features, but weighting was applied to the training data to mimic the magnitude distribution in the target photometric sample. See \cite{kids_annz} for details.

Figure~\ref{fig:comparison} compares the bias and the scatter as a function of photometric redshifts for the $\dense$ sample derived by the red-sequence method versus other approaches in KiDS DR3. We use SMAD of $(z_{\rm phot} - z_{\rm spec})/(1+z_{\rm spec})$ as a proxy for scatter and $\delta z = z_{\rm phot} - z_{\rm spec}$ as bias. Both quantities are computed in bins of photometric redshift. For galaxies in the $\dense$ sample, all four photo-$z$ methods yield nearly the same level of scatter, with the bright ANNz2 approach performing the best (mean scatter of 0.0135), followed by the red-sequence redshifts estimated in this work (with the mean of 0.0152); note however that the former photo-$z$ solution is not available for $z \geq 0.6$. We observe similar trends in the photo-$z$ errors of the $\lum$ sample (not shown).

The high accuracy and precision of the bright ANNz2 redshifts (for $z<0.6$) is not surprising as these photo-$z$'s were specifically trained on the bright sample of GAMA galaxies and were designed to deliver very precise redshift for bright low-redshift galaxies. The red-sequence redshifts estimated with the method in this work are nearly as accurate and precise as the bright ANNz2 redshifts. 
The full-depth ANNz2 redshifts, on the other hand, are highly biased at $z\sim0.4$. This is probably due to the fact that the full-depth ANNz2 sample was trained on the full-depth KiDS DR3 data. As explained in \citet{kids_annz}, the galaxy colours of the spectroscopic sample were re-scaled such that they match the colour distribution of the full-depth KiDS data. This procedure can lead to obtaining more accurate redshifts for a wide range of magnitudes including the deep data at the expense of compromising the ANNz2 photo-$z$ accuracy of bright red galaxies at $z\sim0.4$.  

It is important to note that, although the BPZ redshifts have small scatters for the red galaxies, their bias can be as large as $\delta z \sim 0.04$ for our LRG sample. As discussed in more detail elsewhere \citep{hendrick2017,kids_annz} the BPZ photo-$z$'s available in KiDS DR3 had been optimized for faint higher-redshift galaxies used in cosmic shear analyses. Their worse performance than of machine learning ones which use complete training sets is therefore not surprising. In forthcoming KiDS DR4 this will be much improved in particular thanks to using a prior better optimized for low redshifts. Comparison between the performances of different photo-$z$'s of the red-sequence galaxies is summarized in Table~\ref{tab:comparison}. 

\subsection{Observational systematics}

We assess the robustness of the red-sequence photo-$z$'s against a set of observational systematics. This test is done to ensure that the photometric variations across the survey footprint do not impact the red-sequence photo-$z$ errors. If the photometric redshifts of red galaxies are to be used in large-scale structure and cross correlation studies, they need to have uniform uncertainties across the survey with no strong dependence on photometric variations. 

These observing conditions include the PSF FWHM (measured in arcsec), limiting magnitude (2$\sigma$ in 2 arcsecond apertures), and the 98\% completeness magnitude in the $gri$ filters. 
The first two quantities are derived from the coadded images while the third quantity is derived from the single-band source list. In KiDS DR3, the median value of these quantities is provided for every tile \footnote{\url{http://kids.strw.leidenuniv.nl/DR3/data_table.php}}. There are in total 440 survey tiles over the entire KiDS DR3 footprint. The RA and DEC range of each tile is 62.3 arcmin $\times$ 66.8 arcmin.

We compute the red-sequence photo-$z$ bias and scatter in bins of observing conditions of the survey tiles. We find that the photo-$z$ error distributions are very uniform across different values of the observing conditions. 
This property makes these galaxies an ideal set for clustering studies.

For each observational systematic, we have computed the red-sequence photo-$z$ bias and scatters in the $gri$ bands. We note that the photo-$z$ errors are nearly constant functions of these observational systematics. Figure~\ref{fig:dense_psf} shows the variation of photo-$z$ errors of galaxies in the $\dense$ sample with respect to the (1) PSF FWHM (measured in arcsec) in the $g$ band (right panel), (2) the 2$\sigma$ magnitude limit in the $r$ band (middle panel), and (3) the mean 98\% completeness magnitudes in the $i$ band. In this investigation we limit ourselves to a set of observing conditions that are provided in KiDS DR3. We note however that in future studies employing our LRG samples for cosmological constraints (angular clustering, galaxy-galaxy lensing, etc.) a more in-depth analysis of additional systematics, such as small-scale PSF variations, might be required (\citealt{morrison2015,elvin2017}).

%\begin{figure*}
% \begin{tabular}{rr}
%\includegraphics[width=\textwidth]{figures/jk_north.png} \\
%\includegraphics[width=\textwidth]{figures/jk_south.png}
%\end{tabular}
%\caption{\label{fig:jk} The jackknife resampling regions for estimating the covariance matrix of the galaxy-galaxy lensing signal. The top (bottom) panel shows the jackknife regions in KiDS North (South). The jackknife regions are found using the $kmeans$ algorithm.}
%\end{figure*}

\begin{figure*}

\includegraphics[width=\textwidth]{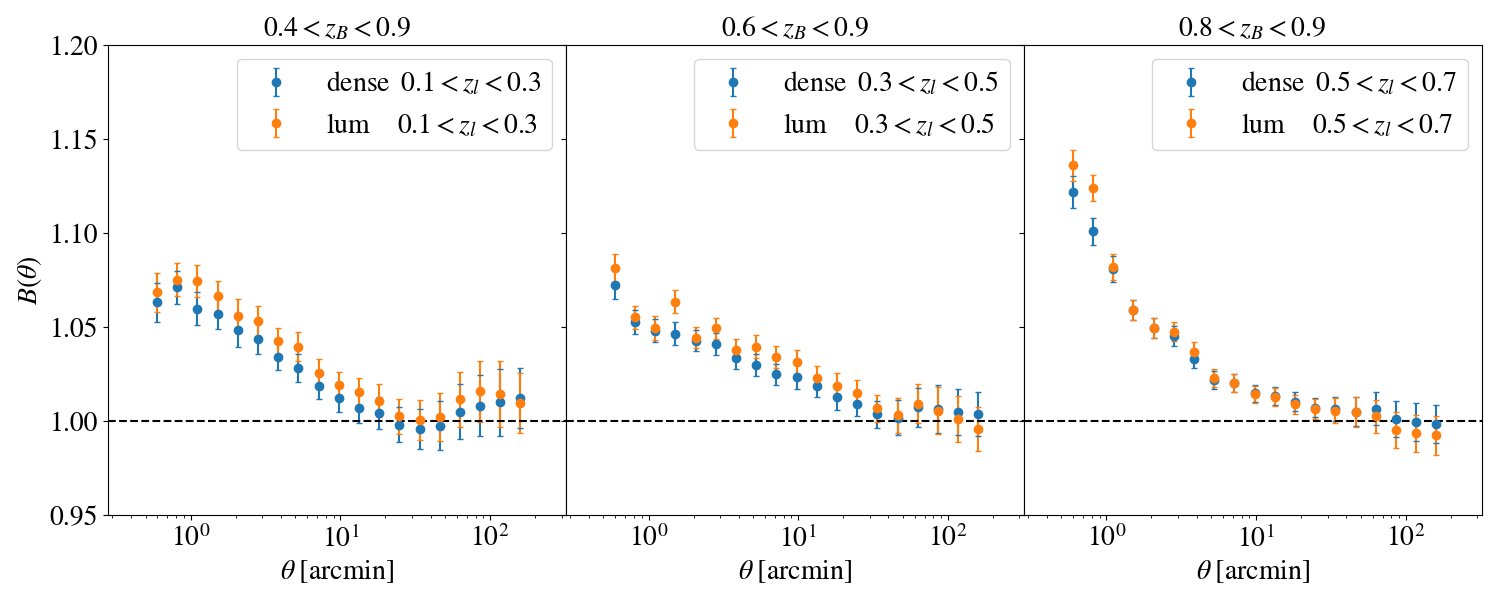}
\caption{\label{fig:boost} Boost factor as a function of angular separation for galaxies in the $\mathtt{dense}$ sample (blue) and those in the $\mathtt{luminous}$ sample (orange). The boost factor is estimated for three redshift bins from left to right: $0.1<z_l<0.3$, $0.3<z_l<0.5$, $0.5<z_l<0.7$. 
We note that on scales larger than $\sim$ 10 arcminute, the estimated boost factors are consistent with one. The errorbars are the square-roots of the diagonal elements of the jackknife error covariance matrices as a function of angular separation. For the first lens redshift bin the boost factor on small scales ($\theta \sim 1 \; \mathrm{arcmin}$) is $\sim$1.05, while for the last two redshift bin the boost factor on small scale can be as large as $\sim 1.1$ and $\sim 1.15$ respectively.}
\end{figure*}

\begin{figure*}
\includegraphics[width=\textwidth]{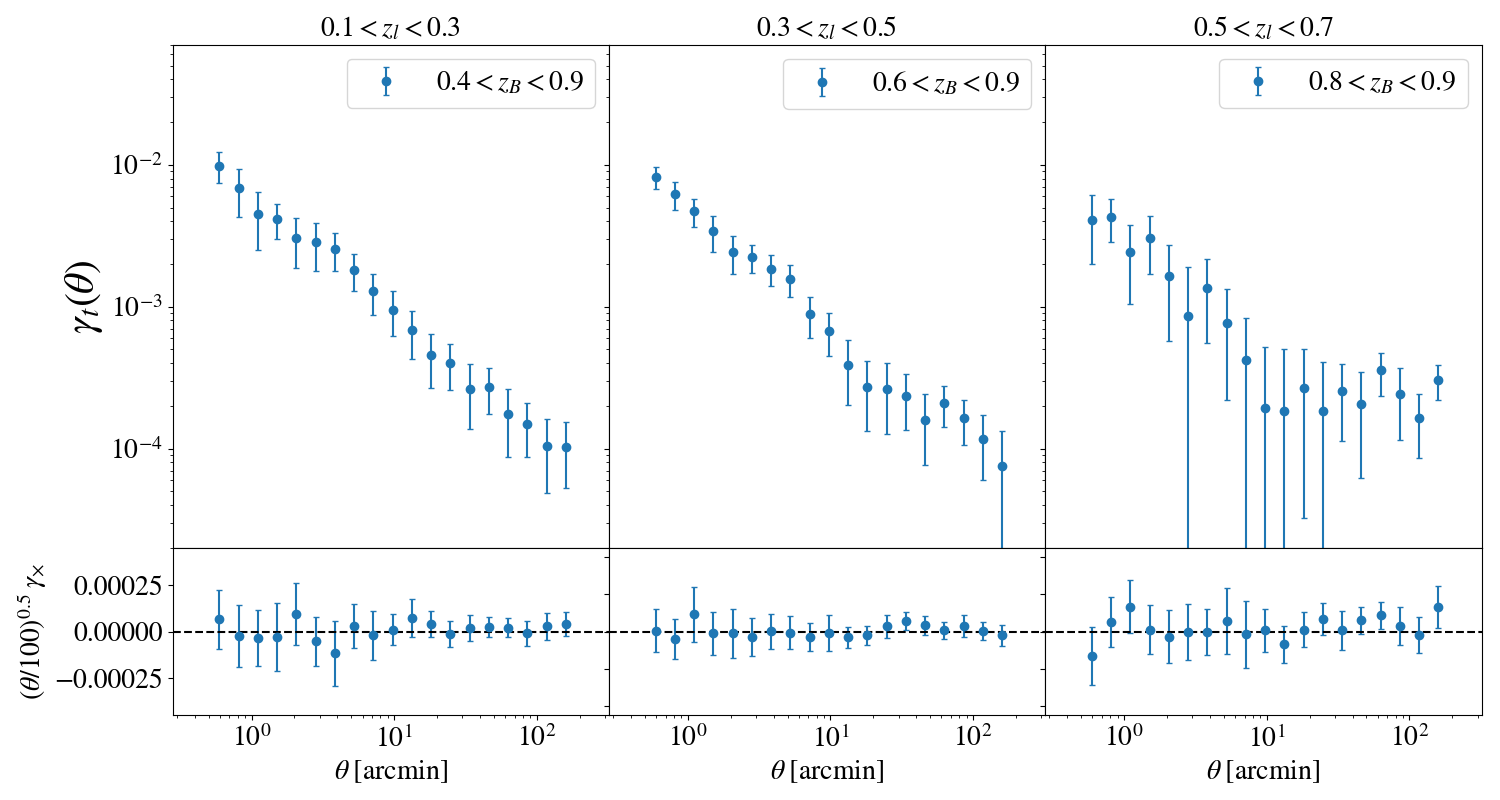} 
\caption{\label{fig:gammat_dense} Tangential shear (top panel) and cross shear (bottom panel) measured around galaxies in the $\mathtt{dense}$ sample for three lens redshift bins from left to right: $0.1<z_l<0.3$, $0.3<z_l<0.5$, $0.5<z_l<0.7$. For each lens redshift bin, sources are selected such that $z_B > \mathrm{max}(z_l)+0.1$. For the shown mean tangential shear signal we have applied the boost factor correction and the random point subtraction, while for the shown mean cross component of shear, we have applied the random subtraction. The uncertainties are derived from the jackknife resampling method. We have scaled the $y$-axis of the bottom panels to make the errorbars on the cross-component more visible. Top panel from left to right: the estimated signal-to-noise ratio of the estimated signal is 20.9, 23.1, and 11.0 respectively. Bottom panel from left to right, the null $\chi^{2}$/ndf for the cross component is 6.3/19, 7.5/19, and 10.0/19 respectively.}
\end{figure*}

\begin{figure*}
\includegraphics[width=\textwidth]{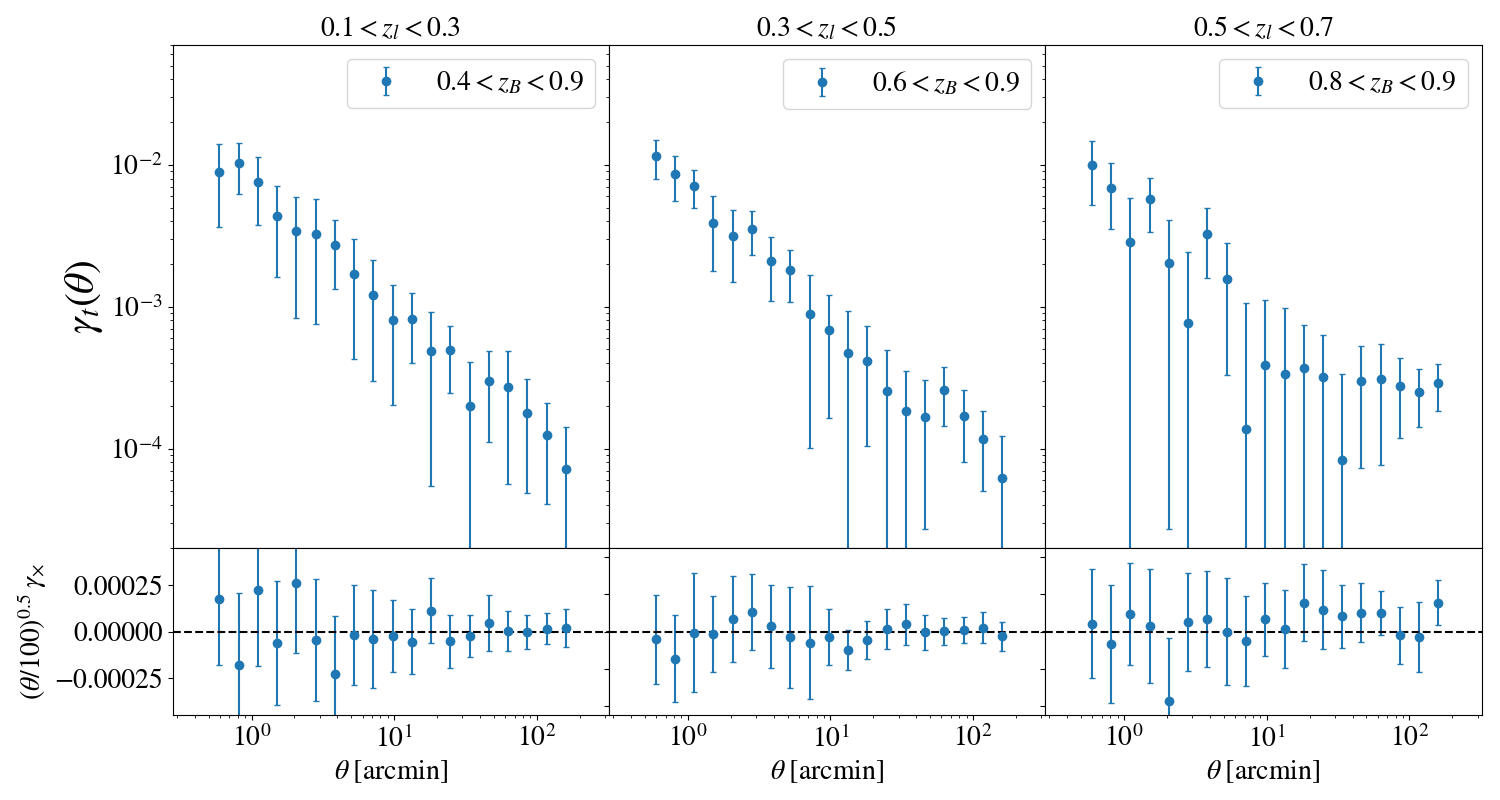}
\caption{\label{fig:gammat_lum} Same as Figure~\ref{fig:gammat_dense} but for galaxies in the $\mathtt{luminous}$ sample. Top panel from left to right: the estimated signal-to-noise ratio of the estimated signal is 11.8, 15.3, and 10.7 respectively. Bottom panel From left to right: the null $\chi^{2}$/ndf for the cross component is 4.7/19, 3.4/19, and 14.1/19 respectively.}
\end{figure*}

%\begin{figure*}
%\includegraphics[width=\textwidth]{new_figures/error.png}
%\caption{\label{fig:error} The impact of random point subtraction on the tangential shear error-bars. Shown are the square-roots of the diagonal elements of the jackknife error covariance matrices as a function of angular separation. The three panels from left to right correspond to errorbars on the tangential shear around the galaxies in the $\mathtt{dense}$ sample in three redshift bins: $0.1<z_l<0.3$, $0.3<z_l<0.5$, $0.5<z_l<0.7$. The blue lines show the derived uncertainties after the random point subtraction and the yellow lines show the derived uncertainties before the random point subtraction. Our results suggest that without the random point subtraction the uncertainties on large scales can be significantly larger.}
%\end{figure*}

\section{Galaxy-galaxy lensing}\label{sec:lensing}

In this section we present lensing measurements around the LRGs using the faint source galaxies in KiDS-450 cosmic shear data \citep{hendrick2017,kids_dr3} by the sample of red galaxies selected in this work. 
We split both lens samples into three tomographic redshift bins of equal widths, in the redshift range of $0.1<z_l<0.7$. 

For each tomographic lens bin, we consider a source bin consisting of galaxies with BPZ redshifts in the redshift range of $\mathrm{max}(z_\mathrm{l})+\delta_z<z_B<0.9$ where $\mathrm{max}(z_\mathrm{l})$ is the maximum redshift of the lens bin under consideration. We choose the value of $\delta_z = 0.1$ in order to maximize the signal-to-noise ratio of the lensing signal while minimizing the contamination of the source population with lens galaxies. Such contamination will dilute the lensing signal particularly on small scales, thus it requires applying a correction to the estimated signal. This correction is also called the \textit{boost factor}, which we will explain shortly. The maximum redshift of sources, $z_B=0.9$, is set by requiring the catastrophic outlier rate to be less than 10\% \citep{kuijken2015}.

\subsection{Cosmic shear data}

We use the cosmic shear measurements presented in \citet{hendrick2017}. Shapes of galaxies are measured by the \textit{lensfit} algorithm (\citealt{lensfit1,lensfit3}; \citealt{lensfit2}), in particular by its most recent implementation in which ellipticities of source galaxies are internally self-calibrated (\citealt{lensfit4}). Source redshifts are estimated with the BPZ algorithm \citep{bpz1999}. Following \citet{hendrick2017}, we only use sources with best-fit photometric redshifts in the range $0.1<z_B<0.9$. Furthermore, most low-redshift and bright sources have been removed by the $m_{\rm r}>20$ cut (\citealt{hendrick2017}). For a more thorough description of the list of criteria for removing flagged source galaxies we refer the reader to \citet{hendrick2017}.

\subsection{Measurements}
We measure the mean tangential shear $\langle \gamma_t \rangle$ and the mean cross-component of the shear $\langle \gamma_{\times}\rangle$. The latter is not produced by gravitational lensing. However, it is a useful test of systematics in the data. Measurement of the tangential and the cross-component of shear is performed in the following way. First, for a pair of lens-source galaxies, the ellipticity of the source galaxy $j$ is decomposed into the tangential and the cross components:
\begin{eqnarray}
e_{t,j} &=& -e_{1,j}\cos(2\phi_j) - e_{2,j}\sin(2\phi_j), \\
e_{\times,j} &=& e_{1,j}\sin(2\phi_j) - e_{2,j}\cos(2\phi_j),
\end{eqnarray}
where $(e_{1,j},e_{2,j})$ are the ellipticity components of the source galaxy $j$ in a Cartesian coordinate system centred on the lens galaxy and $\phi_j$ is the 
position angle of the source galaxy with respect to the horizontal axis in this Cartesian coordinate system. 

Then the mean tangential and cross-components of the shear $\langle \gamma_{t,\times} \rangle$ can be obtained by estimating the mean $\langle e_{t,\times} \rangle$ for a large ensemble of lens-source pairs in the data. Note that this estimator is built under the assumption that the intrinsic galaxy ellipticities are randomly oriented. However, in addition to gravitational lensing, it does receive contribution from the intrinsic alignment of galaxies (for physically close lens-source pairs; see \citealt{blazek2012,clampitt2017}), which needs to be accounted for in the modelling of the signal (\citealt{joudaki2018,edo2018}).  

 We measure $\langle \gamma_{t,\times} \rangle$ using a weighted mean of $e_{t,\times}$:
 
\be
\langle \gamma_{\alpha} (\theta)\rangle = \frac{\sum_{\rm ls}w_{\rm s}e_{\alpha, \mathrm{ls}}}{\sum_{\rm s}w_{\rm s}},
\ee
where $\alpha$ denotes $\{t,\times\}$, the summation is over all pairs of lens ($\mathrm{l}$) source ($\mathrm{s}$) galaxies in an angular bin centred on $\theta$, and $w_{\rm s}$ is the \textit{lensfit} weight assigned to a given source ellipticity. 

Following \citet{viola2015}, in order to account for the multiplicative bias in the cosmic shear data, we apply this correction to the estimated tangential shear: 
\begin{eqnarray}
\langle \gamma_{t} (\theta)\rangle &\rightarrow& \frac{1}{1+\mu(\theta)} \langle \gamma_{t} (\theta)\rangle, \\
\mu &=& \frac{\sum_{\rm ls}w_{\rm s}m_{\rm s}}{\sum_{\rm s}w_{\rm s}},
\end{eqnarray}
where $m_{\rm s}$ is the multiplicative noise bias in the \textit{lensfit} shear estimates (\citealt{lensfit4}). We find that this correction is small and largely independent of the angular separation (see \citealt{viola2015,amon2017a,brouwer2018,dvornik2018} for further discussion of the multiplicative bias correction in KiDS).

We also measure $\langle \gamma_{t} \rangle$ around a set of points randomly distributed across the survey footprint. These random points are generated using the geometry of the survey. In the absence of systematics, such a signal is expected to be zero. In practice however, this signal can be non-negligible due to spatially varying additive shear bias, and the anisotropic distribution of source galaxies around lenses as a result of masked regions and edges of the survey. 
Therefore in order to robustly remove the impact of coherent additive shear bias in the estimated galaxy-galaxy lensing signal, it is important to measure the mean tangential shear around random points and to subtract it from the mean tangential shear around lenses (\citealt{mandelbaum2005,mandelbaum2013,singh2017}). The added advantage of random point subtraction is the decrease of statistical errors on large scales (see also \citealt{prat2017}).
Therefore, we incorporate the random point subtraction into estimation of $\langle \gamma_{\alpha} \rangle$:

\be
\langle \gamma_{\alpha} (\theta)\rangle = \langle \gamma_{\alpha, \mathrm{lens}} (\theta)\rangle - \langle \gamma_{\alpha, \mathrm{random}} (\theta)\rangle.
\label{eq:lens-random}
\ee

Excess source counts around the lenses can bias our estimate of the tangential shear. Any sources that in fact are associated with the lenses would not be lensed, resulting in suppression of the lensing signal at small angular separations\footnote{We have implicitly assumed that the contribution from intrinsic alignment of physically close source-lens pairs is negligible.}.
We correct this effect by applying the so-called boost correction to the estimated tangential shear.  
We estimate the boost correction by computing the excess of sources around lenses compared to the random points. We define a boost factor parameter $B(\theta)$ in the following way:

\be
B(\theta) = \frac{N_{\mathrm{random}}}{N_{\mathrm{lens}}}\frac{\sum_{\mathrm{l,s}}w_{\rm ls}}{\sum_{\mathrm{r,s}}w_{\rm rs}},
\label{eq:boost_def}
\ee
where $N_{\mathrm{random}}$ ($N_{\mathrm{lens}}$) denotes the number of randoms (lenses), $w_{\rm ls}$ ($w_{\rm rs}$) is the weight assigned to the lens-source (random-source) pair in the angular bin centred on $\theta$, and the summation in the numerator (denominator) is over all the lens-source (random-source) pairs in the data. 
The boost correction is expected to be close to unity on large  angular scales but it can be significant (as large as 10\%) on very small scales. Furthermore, since the photo-$z$ uncertainties of source galaxies increase with redshift, the excess counts increase when considering the high redshift source bins. 
Finally, we modify the estimator of the tangential shear (\ref{eq:lens-random}) in the following way:

\be
\langle \gamma_{t} (\theta)\rangle = B(\theta)\big(\langle \gamma_{t, \mathrm{lens}} (\theta)\rangle - \langle \gamma_{t, \mathrm{random}} (\theta)\rangle \big)
\label{eq:boost_cor}
\ee

The galaxy-galaxy lensing measurements presented in this work are 
measured in 19 logarithmically spaced angular bins between 0.5 and 250 arcmin. The lensing measurements are not extended to smaller scales as small-scale lensing may suffer from blending of source galaxies from the deep imaging data with the foreground lenses that are typically much brighter than the source galaxies. Furthermore, including the smaller scales could result in a lack of source galaxies that are behind (or in angular vicinity of) foreground lenses. Such obscuration, unlike the physically associated source galaxies, can result in a boost factor that is smaller than one. These issues can be avoided by a conservative cut on angular scales at 0.5 arcmin. All the measurements are computed using the $\mathtt{TreeCorr}$ software\footnote{\url{https://github.com/rmjarvis/TreeCorr}}.

%\begin{figure}
% \begin{tabular}{cc}
%\includegraphics[width=0.5\textwidth]{figures/gamma_psf.png}
%\includegraphics[width=0.5\textwidth]{figures/gamma_random.png}
%\end{tabular}
%\caption{\label{fig:psf_random} Tangential shear around the random points in three redshift bins: $0.4<z_B<0.9$ (blue), $0.6<z_B<0.9$ (yellow), $0.8<z_l<0.9$ (green). By evaluating the null $\chi^{2}$ we confirm that this signal is consistent with zero.}
%\end{figure}

\subsection{Covariance estimation}
We estimate the measurement uncertainties using the jackknife resampling method (\citealt{norberg2009,oliver2016,singh2017,shirasaki2017}). 
In the jackknife method, the survey footprint is first divided into $N_{\mathrm{JK}}$ jackknife subregions of approximately equal area. 
Then for each subregion $k\in\{1,...,N_{\mathrm{JK}}\}$, the lensing data vector $\boldsymbol{\gamma}_{\alpha}^{(k)} = \langle\gamma^{(k)}_{\alpha}\rangle$ is measured by cutting out the $k$-th subregion and estimating the lensing signal of the rest of the survey footprint. Note that $\boldsymbol{\gamma}_{\alpha}^{(k)}$ is a 19-dimensional vector which contains the tangential (cross) component of the shear in all angular bins. The jackknife estimator of the covariance matrix is then given by:
\be 
C_{\rm JK, \alpha} = \frac{\njk - 1}{\njk}\sum_{k=1}^{\njk}\big(\dk-\dbar\big)^{T}\big(\dk-\dbar\big), 
\label{eq:jk}
\ee
where $\dbar$ is the mean of all $\boldsymbol{\gamma}_{\alpha}^{(k)}$ vectors. 

In order to construct jackknife subregions, we generate a large number of random points uniformly distributed across the entire KiDS DR3 footprint. Then we use the kmeans\footnote{\url{https://github.com/esheldon/kmeans_radec}} algorithm to divide the random points into 100 disjoint subregions each encompassing a nearly equal number of random points. Note that for the purpose of determining the jackknife subregions, we exclude the small disjoint regions of the KiDS DR3 footprint that are between the G9, G12, G15, G23, and GS patches. 

Finally, we compute the unbiased estimate of the inverse covariance matrix by applying the correction (\citealt{hartlap2007}):
\be
\widehat{C^{-1}} = \frac{N_{\rm JK}-N_{\rm bins}-2}{N_{\rm JK}-1}\widehat{C}^{-1},
\label{eq:jk_inv}
\ee
where $\widehat{C}$ is the jackknife estimate of the covariance matrix and is given by Eq.~\ref{eq:jk}, and $N_{\rm bins}$ is the number of angular bins. 

\subsection{Results}

The estimated boost corrections are demonstrated in Fig.~\ref{fig:boost}. The blue (orange) points show the boost factor applied to $\langle \gamma_t \rangle$ measurements for the $\dense$ ($\lum$) sample. The errorbars are derived from the jackknife resampling method. As expected, the boost corrections deviate from one at small angular scales and are consistent with one at large angular scales. $B(\theta)$ is larger for higher redshifts because the probability of physical association of sources with lens galaxies is higher. This highlights the importance of accounting for the effect of physical association of sources with lenses.

Our estimates of $\langle \gamma_{t\times}\rangle$ for lenses in the $\dense$ and the $\lum$ samples 
are shown in Figs.~\ref{fig:gammat_dense} and~\ref{fig:gammat_lum} respectively. The errorbars are derived by taking the square root of the diagonal elements of the jackknife covariance matrix for each observable. We note that in a given tomographic lens redshift bin, the estimated uncertainties of $\langle \gamma_t\rangle$ measured for galaxies in the $\lum$ sample are larger than those of $\langle \gamma_t\rangle$ measured for the $\dense$ bin. This is due to the fact that by construction at a given redshift, the comoving number density of the $\lum$ galaxies is much smaller the that of the $\dense$ galaxies resulting in fewer lens-source pairs at any angular bin. 

We also note that for both samples of lens galaxies, $\langle \gamma_t\rangle$ is the noisiest for the last tomographic lens redshift bin: $0.5<z_l<0.7$. The corresponding tomographic source bin of this lens redshift bin  contains galaxies with only $0.8<z_B<0.9$ which yields a very limited number of lens-source pairs. In order to assess the detection significance of the measurements, we compute the signal-to-noise ratios defined in the following way:
\be 
S/N = \sqrt{\dg^{T} C^{-1}\dg},
\label{eq:snr}
\ee
where $C^{-1}$ is the estimate of the inverse covariance matrix (Eq.~\ref{eq:jk_inv}), and $\dg$ denotes the measured tangential shear in a tomographic lens bin for a lens sample. Equation~\ref{eq:snr} can be interpreted as the ratio of the mean and the square-root of the variance of the probability distribution function that the measurements are drawn from\footnote{In principle, $\dg$ in~\ref{eq:snr} is provided by a theoretical model. Since we only present the measurements and we postpone the modelling to future analyses, we use the measurements to obtain an approximate $S/N$.}. For galaxies in the $\dense$ sample the signal-to-noise ratios of the detected tangential signals are 20.9, 23.1, and 11.0 in the tomographic lens bins $0.1<z_l<0.3$, $0.3<z_l<0.5$, and $0.5<z_l<0.7$ respectively. For galaxies in the $\lum$ sample the signal-to-noise ratios of the detected tangential signals are 11.8, 15.3, and 10.7 in the tomographic lens bins $0.1<z_l<0.3$, $0.3<z_l<0.5$, and $0.5<z_l<0.7$ respectively.
Note that the top panels of Figs~\ref{fig:gammat_dense},~\ref{fig:gammat_lum} show the tangential shear after random point subtraction (Eq.~\ref{eq:lens-random}) and boost correction (see Eqs.~\ref{eq:boost_def},~\ref{eq:boost_cor}).

Additionally, we present the cross-component measurements in the bottom panels of 
Figs.~\ref{fig:gammat_dense},~\ref{fig:gammat_lum}. These signals are shown in the bottom panels of Figs.~\ref{fig:gammat_dense},~\ref{fig:gammat_lum} in which the $y$-axis has been scaled  for better visibility of the errorbars. 

We compute the Null $\chi^{2}$ per number of degrees of freedom for the following null data vectors: cross component of shear $\langle \gamma_{\times}\rangle$ and the tangential shear around randoms $\langle \gamma_{t, \mathrm{random}} (\theta)\rangle$.
For a given null data vector $\mathbf{x}_{\rm Null}$, and the inverse covariance matrix associated with the null signal by $C^{-1}$, the Null $\chi^{2}$ is given by:
\be
\chi^{2} = \mathbf{x}^{T}_{\rm Null}C^{-1}\mathbf{x}_{\rm Null}.
\ee

In order for a measurement to pass a null test, $\chi^{2}/\mathrm{ndf}$ needs to be smaller than or equal to one. In order of the lens redshift bin, the measured Null $\chi^{2}/\mathrm{ndf}$ are 6.3/19, 7.5/19, 10.0/19 for lenses in the $\dense$ sample and 4.7/19, 3.4/19, 14.1/19 for lenses in the $\lum$ sample. For the mean tangential shear around randoms, in order of tomographic bins, the null $\chi^{2}\mathrm{ndf}$ are 6.5/19, 4.5/19, and 1.1/19. This implies that the random shear signal is consistent with zero.

\section{Summary and conclusion}\label{sec:summary}

In this investigation we have presented the selection and weak lensing analysis of luminous red galaxies with the Kilo-Degree Survey broadband photometry. We exploited the KiDS multi-band imaging data and the overlapping spectroscopic datasets to select two samples of red galaxies with different luminosity thresholds and comoving densities. Since these galaxies are mostly bright, they are complementary to the fainter galaxies that are used for cosmic shear studies. As a result, the selection of these galaxies is a crucial step towards fully realizing the scientific potential of the Kilo-Degree Survey.  

We have shown that these galaxies have very accurate and precise redshifts. The estimated red-sequence photometric redshifts of these galaxies are nearly as accurate and precise as the redshifts obtained by the ANNz2 algorithm trained on a complete sample of bright galaxies in the GAMA survey. A nice property of these red galaxies is that regardless of the photo-$z$ estimation method, they have nearly equal photo-$z$ scatters, although we note that the bright ANNz2 photo-$z$s are the most stable in terms of bias.
%However, the least biased photo-$z$'s of these galaxies are provided by the red-sequence and the bright ANNz2 redshifts.

We have also demonstrated that the estimated LRG redshifts are very robust against a number of survey observing conditions. These conditions include the seeing, limiting $gri$ magnitudes, and the 98\% completeness magnitudes of the survey tiles. The photometric redshift uncertainties are uniform and do not vary with photometric variations across survey tiles. These qualities make these red-sequence galaxies and their estimated red-sequence redshifts an ideal dataset for galaxy clustering and cross-correlation studies. As an example of the scientific application of this sample of galaxies, we have presented galaxy-galaxy lensing measurements. Using the KiDS shear data, we have found a significant detection of tangential shear even for LRGs with $0.5<z_l<0.7$. 

The longest wavelength used in this work for the red-sequence selection was covered by the $i$ band. This limited the redshift range of identified LRGs to $z<0.7$. In order to extend the method to higher redshifts, one needs to use additionally near-infrared (NIR) bands such as $Z$ \citep{redmap_sdss,redmap_des}. This is indeed possible by combining optical KiDS data with the VISTA  Kilo degree  INfrared  Galaxy  survey (VIKING; \citealt{viking2013}),
%Furthermore, applying the method described in this paper to the combination of the optical and the near infrared (NIR) photometry can significantly improve the selection of red galaxies. Combination of the Kilo Degree Survey, probing the optical wavelengths, and the VISTA  Kilo degree  INfrared  Galaxy  survey (VIKING; \citealt{viking2013,viking2015})  survey,
probing  the  NIR  wavelengths (8000-24000 {{{\AA}}}).%, enables production of one of 
This will provide the largest existing joint optical-NIR dataset %(\citealt{KV450}) 
for cosmological studies. We will present the selection of bright red-sequence galaxies from the joint optical-NIR catalogue in a future work.

Weak lensing analysis of the upcoming 1000 deg$^2$ photometry of the Kilo-Degree Survey will provide tight constraints on cosmological parameters. Selection of a set of red-sequence galaxies with reliable redshifts with the 1000 deg$^2$ photometry will enable us to measure additional probes of the large-scale structure, such as galaxy clustering and galaxy-galaxy lensing. Joint analysis of these additional probes and the cosmic shear will help improve the constraints on cosmological models. Additionally, the red-sequence galaxy catalogue will provide a useful playground for testing the empirical models of galaxy-halo connection and the intrinsic alignments of galaxies.

\section*{Acknowledgements}
We thank Thomas Erbens for reading the manuscript 
and providing valuable feedbacks.
MV and HHo acknowledge support from Vici grant 639.043.512 
from the Netherlands Organization of Scientific Research (NWO).
MB is supported by the NWO through grant number 614.001.451.

This research is based on data products from observations
made with ESO Telescopes at the La Silla Paranal
Observatory under programme IDs 177.A-3016, 177.A-3017
and 177.A-3018, and on data products produced by Target
OmegaCEN, INAF-OACN, INAF-OAPD and the KiDS
production team, on behalf of the KiDS consortium. OmegaCEN
and the KiDS production team acknowledge support
by NOVA and NWO-M grants. Members of INAF-OAPD
and INAF-OACN also acknowledge the support from the
Department of Physics \& Astronomy of the University of
Padova, and of the Department of Physics of Univ. Federico
II (Naples).

GAMA is a joint European-Australasian project
based around a spectroscopic campaign using the AngloAustralian
Telescope. The GAMA input catalogue is based
on data taken from the Sloan Digital Sky Survey and the
UKIRT Infrared Deep Sky Survey. Complementary imaging
of the GAMA regions is being obtained by a number of independent
survey programs including GALEX MIS, VST
KiDS, VISTA VIKING, WISE, Herschel-ATLAS, GMRT
and ASKAP providing UV to radio coverage. GAMA is
funded by the STFC (UK), the ARC (Australia), the AAO,
and the participating institutions. The GAMA website is
\url{www.gama-survey.org}.

2dFLenS is based on data acquired through the Australian Astronomical Observatory, under programme A/2014B/008.  It would not have been possible without the dedicated work of the staff of the AAO in the development and support of the 2dF-AAOmega system, and the running of the AAT.

Funding for SDSS-III was provided by the Alfred P. Sloan Foundation, the Participating Institutions, the National Science Foundation, and the U.S. Department of Energy Office of Science. The SDSS-III website is \url{http://www.sdss3.org/}. SDSS-III is managed by the Astrophysical Research Consortium for the Participating Institutions of the SDSS-III Collaboration including the University of Arizona, the Brazilian Participation Group, Brookhaven National Laboratory, Carnegie Mellon University, University of Florida, the French Participation Group, the German Participation Group, Harvard University, the Instituto de Astrofisica de Canarias, the Michigan State/Notre Dame/JINA Participation Group, Johns Hopkins University, Lawrence Berkeley National Laboratory, Max Planck Institute for Astrophysics, Max Planck Institute for Extraterrestrial Physics, New Mexico State University, New York University, Ohio State University, Pennsylvania State University, University of Portsmouth, Princeton University, the Spanish Participation Group, University of Tokyo, University of Utah, Vanderbilt University, University of Virginia, University of Washington, and Yale University.

This work has made use of python (\url{www.python.org}), including the packages numpy (\url{www.numpy.org}), scipy
(\url{www.scipy.org}). Plots have been produced with matplotlib.

%%%%%%%%%%%%%%%%%%%%%%%%%%%%%%%%%%%%%%%%%%%%%%%%%%

% BibTeX:

\bibliographystyle{mnras}
\bibliography{references}

\begin{thebibliography}{}
\makeatletter
\relax
\def\mn@urlcharsother{\let\do\@makeother \do\$\do\&\do\#\do\^\do\_\do\%\do\~}
\def\mn@doi{\begingroup\mn@urlcharsother \@ifnextchar [ {\mn@doi@}
  {\mn@doi@[]}}
\def\mn@doi@[#1]#2{\def\@tempa{#1}\ifx\@tempa\@empty \href
  {http://dx.doi.org/#2} {doi:#2}\else \href {http://dx.doi.org/#2} {#1}\fi
  \endgroup}
\def\mn@eprint#1#2{\mn@eprint@#1:#2::\@nil}
\def\mn@eprint@arXiv#1{\href {http://arxiv.org/abs/#1} {{\tt arXiv:#1}}}
\def\mn@eprint@dblp#1{\href {http://dblp.uni-trier.de/rec/bibtex/#1.xml}
  {dblp:#1}}
\def\mn@eprint@#1:#2:#3:#4\@nil{\def\@tempa {#1}\def\@tempb {#2}\def\@tempc
  {#3}\ifx \@tempc \@empty \let \@tempc \@tempb \let \@tempb \@tempa \fi \ifx
  \@tempb \@empty \def\@tempb {arXiv}\fi \@ifundefined
  {mn@eprint@\@tempb}{\@tempb:\@tempc}{\expandafter \expandafter \csname
  mn@eprint@\@tempb\endcsname \expandafter{\@tempc}}}

\bibitem[\protect\citeauthoryear{{Albareti} et~al.,}{{Albareti}
  et~al.}{2017}]{sdss_dr13}
{Albareti} F.~D.,  et~al., 2017, \mn@doi [\apjs] {10.3847/1538-4365/aa8992},
  \href {http://adsabs.harvard.edu/abs/2017ApJS..233...25A} {233, 25}

\bibitem[\protect\citeauthoryear{{Amon} et~al.,}{{Amon}
  et~al.}{2017b}]{amon2017a}
{Amon} A.,  et~al., 2017b, preprint, \href
  {http://adsabs.harvard.edu/abs/2017arXiv170704105A} {} (\mn@eprint {arXiv}
  {1707.04105})

\bibitem[\protect\citeauthoryear{{Amon} et~al.,}{{Amon}
  et~al.}{2017a}]{amon2017}
{Amon} A.,  et~al., 2017a, preprint, \href
  {http://adsabs.harvard.edu/abs/2017arXiv171110999A} {} (\mn@eprint {arXiv}
  {1711.10999})

\bibitem[\protect\citeauthoryear{{Amon} et~al.,}{{Amon}
  et~al.}{2018}]{amon2018}
{Amon} A.,  et~al., 2018, \mn@doi [\mnras] {10.1093/mnras/sty859}, \href
  {http://adsabs.harvard.edu/abs/2018MNRAS.tmp..834A} {}

\bibitem[\protect\citeauthoryear{{Ben{\'{\i}}tez}}{{Ben{\'{\i}}tez}}{2000}]{bpz1999}
{Ben{\'{\i}}tez} N.,  2000, \mn@doi [\apj] {10.1086/308947}, \href
  {http://adsabs.harvard.edu/abs/2000ApJ...536..571B} {536, 571}

\bibitem[\protect\citeauthoryear{{Bertin} \& {Arnouts}}{{Bertin} \&
  {Arnouts}}{1996}]{sextractor}
{Bertin} E.,  {Arnouts} S.,  1996, \mn@doi [\aaps] {10.1051/aas:1996164}, \href
  {http://adsabs.harvard.edu/abs/1996A%26AS..117..393B} {117, 393}

\bibitem[\protect\citeauthoryear{{Bilicki} et~al.,}{{Bilicki}
  et~al.}{2018}]{kids_annz}
{Bilicki} M.,  et~al., 2018, \mn@doi [\aap] {10.1051/0004-6361/201731942},
  \href {http://adsabs.harvard.edu/abs/2018A%26A...616A..69B} {616, A69}

\bibitem[\protect\citeauthoryear{{Blake} et~al.,}{{Blake}
  et~al.}{2016}]{blake2016}
{Blake} C.,  et~al., 2016, \mn@doi [\mnras] {10.1093/mnras/stw1990}, \href
  {http://adsabs.harvard.edu/abs/2016MNRAS.462.4240B} {462, 4240}

\bibitem[\protect\citeauthoryear{{Blazek}, {Mandelbaum}, {Seljak}  \&
  {Nakajima}}{{Blazek} et~al.}{2012}]{blazek2012}
{Blazek} J.,  {Mandelbaum} R.,  {Seljak} U.,   {Nakajima} R.,  2012, \mn@doi
  [Journal of Cosmology and Astro-Particle Physics]
  {10.1088/1475-7516/2012/05/041}, \href
  {https://ui.adsabs.harvard.edu/#abs/2012JCAP...05..041B} {2012, 041}

\bibitem[\protect\citeauthoryear{{Bolzonella}, {Miralles}  \&
  {Pell{\'o}}}{{Bolzonella} et~al.}{2000}]{Bolzonella2000}
{Bolzonella} M.,  {Miralles} J.-M.,   {Pell{\'o}} R.,  2000, \aap, \href
  {http://adsabs.harvard.edu/abs/2000A%26A...363..476B} {363, 476}

\bibitem[\protect\citeauthoryear{{Bovy}, {Hogg}  \& {Roweis}}{{Bovy}
  et~al.}{2010}]{xd_code}
{Bovy} J.,  {Hogg} D.~W.,   {Roweis} S.~T.,  2010, {Extreme Deconvolution:
  Density Estimation using Gaussian Mixtures in the Presence of Noisy,
  Heterogeneous and Incomplete Data}, Astrophysics Source Code Library
  (\mn@eprint {ascl} {1010.032})

\bibitem[\protect\citeauthoryear{{Bovy}, {Hogg}  \& {Roweis}}{{Bovy}
  et~al.}{2011}]{xd_paper}
{Bovy} J.,  {Hogg} D.~W.,   {Roweis} S.~T.,  2011, \mn@doi [Annals of Applied
  Statistics] {10.1214/10-AOAS439}, \href
  {http://adsabs.harvard.edu/abs/2011AnApS...5.1657B} {5}

\bibitem[\protect\citeauthoryear{{Brammer}, {van Dokkum}  \& {Coppi}}{{Brammer}
  et~al.}{2008}]{Brammer2008}
{Brammer} G.~B.,  {van Dokkum} P.~G.,   {Coppi} P.,  2008, \mn@doi [\apj]
  {10.1086/591786}, \href {http://adsabs.harvard.edu/abs/2008ApJ...686.1503B}
  {686, 1503}

\bibitem[\protect\citeauthoryear{{Brouwer} et~al.,}{{Brouwer}
  et~al.}{2018}]{brouwer2018}
{Brouwer} M.~M.,  et~al., 2018, preprint, \href
  {http://adsabs.harvard.edu/abs/2018arXiv180500562B} {} (\mn@eprint {arXiv}
  {1805.00562})

\bibitem[\protect\citeauthoryear{{Bruzual} \& {Charlot}}{{Bruzual} \&
  {Charlot}}{2003}]{bc03}
{Bruzual} G.,  {Charlot} S.,  2003, \mn@doi [\mnras]
  {10.1046/j.1365-8711.2003.06897.x}, \href
  {http://adsabs.harvard.edu/abs/2003MNRAS.344.1000B} {344, 1000}

\bibitem[\protect\citeauthoryear{{Byrd}, {Nocedal}  \& {Schnabel}}{{Byrd}
  et~al.}{1994}]{bfgs}
{Byrd} R.~H.,  {Nocedal} J.,   {Schnabel} R.~B.,  1994, \mn@doi [Mathematical
  Programming] {10.1007/BF01582063}, 63, 129

\bibitem[\protect\citeauthoryear{{Cacciato}, {van den Bosch}, {More}, {Mo}  \&
  {Yang}}{{Cacciato} et~al.}{2013}]{cacciato2013}
{Cacciato} M.,  {van den Bosch} F.~C.,  {More} S.,  {Mo} H.,   {Yang} X.,
  2013, \mn@doi [\mnras] {10.1093/mnras/sts525}, \href
  {http://adsabs.harvard.edu/abs/2013MNRAS.430..767C} {430, 767}

\bibitem[\protect\citeauthoryear{{Capaccioli} et~al.,}{{Capaccioli}
  et~al.}{2012}]{vst}
{Capaccioli} M.,  et~al., 2012, in Science from the Next Generation Imaging and
  Spectroscopic Surveys. p.~1

\bibitem[\protect\citeauthoryear{{Cawthon} et~al.,}{{Cawthon}
  et~al.}{2017}]{cawthon2017}
{Cawthon} R.,  et~al., 2017, preprint, \href
  {http://adsabs.harvard.edu/abs/2017arXiv171207298C} {} (\mn@eprint {arXiv}
  {1712.07298})

\bibitem[\protect\citeauthoryear{{Chabrier}}{{Chabrier}}{2003}]{chabrier2003}
{Chabrier} G.,  2003, \mn@doi [\pasp] {10.1086/376392}, \href
  {http://adsabs.harvard.edu/abs/2003PASP..115..763C} {115, 763}

\bibitem[\protect\citeauthoryear{{Clampitt} et~al.,}{{Clampitt}
  et~al.}{2017}]{clampitt2017}
{Clampitt} J.,  et~al., 2017, \mn@doi [\mnras] {10.1093/mnras/stw2988}, \href
  {http://adsabs.harvard.edu/abs/2017MNRAS.465.4204C} {465, 4204}

\bibitem[\protect\citeauthoryear{{Davis} et~al.,}{{Davis}
  et~al.}{2017}]{davis2017}
{Davis} C.,  et~al., 2017, preprint, \href
  {http://adsabs.harvard.edu/abs/2017arXiv171002517D} {} (\mn@eprint {arXiv}
  {1710.02517})

\bibitem[\protect\citeauthoryear{{Dawson} et~al.,}{{Dawson}
  et~al.}{2013}]{dawson2013}
{Dawson} K.~S.,  et~al., 2013, \mn@doi [\aj] {10.1088/0004-6256/145/1/10},
  \href {http://adsabs.harvard.edu/abs/2013AJ....145...10D} {145, 10}

\bibitem[\protect\citeauthoryear{{Dawson} et~al.,}{{Dawson}
  et~al.}{2016}]{dawson2016}
{Dawson} K.~S.,  et~al., 2016, \mn@doi [\aj] {10.3847/0004-6256/151/2/44},
  \href {http://adsabs.harvard.edu/abs/2016AJ....151...44D} {151, 44}

\bibitem[\protect\citeauthoryear{{Driver} et~al.,}{{Driver}
  et~al.}{2011}]{driver2011}
{Driver} S.~P.,  et~al., 2011, \mn@doi [\mnras]
  {10.1111/j.1365-2966.2010.18188.x}, \href
  {http://adsabs.harvard.edu/abs/2011MNRAS.413..971D} {413, 971}

\bibitem[\protect\citeauthoryear{{Duncan}, {Jarvis}, {Brown}  \&
  {R{\"o}ttgering}}{{Duncan} et~al.}{2018}]{Duncan2018}
{Duncan} K.~J.,  {Jarvis} M.~J.,  {Brown} M.~J.~I.,   {R{\"o}ttgering}
  H.~J.~A.,  2018, \mn@doi [\mnras] {10.1093/mnras/sty940}, \href
  {http://adsabs.harvard.edu/abs/2018MNRAS.477.5177D} {477, 5177}

\bibitem[\protect\citeauthoryear{{Dvornik} et~al.,}{{Dvornik}
  et~al.}{2018}]{dvornik2018}
{Dvornik} A.,  et~al., 2018, preprint, \href
  {http://adsabs.harvard.edu/abs/2018arXiv180200734D} {} (\mn@eprint {arXiv}
  {1802.00734})

\bibitem[\protect\citeauthoryear{{Edge}, {Sutherland}, {Kuijken}, {Driver},
  {McMahon}, {Eales}  \& {Emerson}}{{Edge} et~al.}{2013}]{viking2013}
{Edge} A.,  {Sutherland} W.,  {Kuijken} K.,  {Driver} S.,  {McMahon} R.,
  {Eales} S.,   {Emerson} J.~P.,  2013, The Messenger, \href
  {http://adsabs.harvard.edu/abs/2013Msngr.154...32E} {154, 32}

\bibitem[\protect\citeauthoryear{{Elvin-Poole} et~al.,}{{Elvin-Poole}
  et~al.}{2017}]{elvin2017}
{Elvin-Poole} J.,  et~al., 2017, preprint, \href
  {http://adsabs.harvard.edu/abs/2017arXiv170801536E} {} (\mn@eprint {arXiv}
  {1708.01536})

\bibitem[\protect\citeauthoryear{{Feldmann} et~al.,}{{Feldmann}
  et~al.}{2006}]{Feldmann2006}
{Feldmann} R.,  et~al., 2006, \mn@doi [\mnras]
  {10.1111/j.1365-2966.2006.10930.x}, \href
  {http://adsabs.harvard.edu/abs/2006MNRAS.372..565F} {372, 565}

\bibitem[\protect\citeauthoryear{{Fenech Conti}, {Herbonnet}, {Hoekstra},
  {Merten}, {Miller}  \& {Viola}}{{Fenech Conti} et~al.}{2017}]{lensfit4}
{Fenech Conti} I.,  {Herbonnet} R.,  {Hoekstra} H.,  {Merten} J.,  {Miller} L.,
    {Viola} M.,  2017, \mn@doi [\mnras] {10.1093/mnras/stx200}, \href
  {http://adsabs.harvard.edu/abs/2017MNRAS.467.1627F} {467, 1627}

\bibitem[\protect\citeauthoryear{{Firth}, {Lahav}  \& {Somerville}}{{Firth}
  et~al.}{2003}]{Firth2003}
{Firth} A.~E.,  {Lahav} O.,   {Somerville} R.~S.,  2003, \mn@doi [\mnras]
  {10.1046/j.1365-8711.2003.06271.x}, \href
  {http://adsabs.harvard.edu/abs/2003MNRAS.339.1195F} {339, 1195}

\bibitem[\protect\citeauthoryear{{Friedrich}, {Seitz}, {Eifler}  \&
  {Gruen}}{{Friedrich} et~al.}{2016}]{oliver2016}
{Friedrich} O.,  {Seitz} S.,  {Eifler} T.~F.,   {Gruen} D.,  2016, \mn@doi
  [\mnras] {10.1093/mnras/stv2833}, \href
  {http://adsabs.harvard.edu/abs/2016MNRAS.456.2662F} {456, 2662}

\bibitem[\protect\citeauthoryear{{Gerdes}, {Sypniewski}, {McKay}, {Hao},
  {Weis}, {Wechsler}  \& {Busha}}{{Gerdes} et~al.}{2010}]{Gerdes2010}
{Gerdes} D.~W.,  {Sypniewski} A.~J.,  {McKay} T.~A.,  {Hao} J.,  {Weis} M.~R.,
  {Wechsler} R.~H.,   {Busha} M.~T.,  2010, \mn@doi [\apj]
  {10.1088/0004-637X/715/2/823}, \href
  {http://adsabs.harvard.edu/abs/2010ApJ...715..823G} {715, 823}

\bibitem[\protect\citeauthoryear{{Gladders} \& {Yee}}{{Gladders} \&
  {Yee}}{2000}]{gladders_yee2000}
{Gladders} M.~D.,  {Yee} H.~K.~C.,  2000, \mn@doi [\aj] {10.1086/301557}, \href
  {http://adsabs.harvard.edu/abs/2000AJ....120.2148G} {120, 2148}

\bibitem[\protect\citeauthoryear{{Gunn} et~al.,}{{Gunn}
  et~al.}{2006}]{gunn2006}
{Gunn} J.~E.,  et~al., 2006, \mn@doi [\aj] {10.1086/500975}, \href
  {http://adsabs.harvard.edu/abs/2006AJ....131.2332G} {131, 2332}

\bibitem[\protect\citeauthoryear{{Hao} et~al.,}{{Hao} et~al.}{2009}]{hao2009}
{Hao} J.,  et~al., 2009, \mn@doi [\apj] {10.1088/0004-637X/702/1/745}, \href
  {http://adsabs.harvard.edu/abs/2009ApJ...702..745H} {702, 745}

\bibitem[\protect\citeauthoryear{{Hartlap}, {Simon}  \& {Schneider}}{{Hartlap}
  et~al.}{2007}]{hartlap2007}
{Hartlap} J.,  {Simon} P.,   {Schneider} P.,  2007, \mn@doi [\aap]
  {10.1051/0004-6361:20066170}, \href
  {http://adsabs.harvard.edu/abs/2007A%26A...464..399H} {464, 399}

\bibitem[\protect\citeauthoryear{{Heymans} et~al.,}{{Heymans}
  et~al.}{2013}]{heymans2013}
{Heymans} C.,  et~al., 2013, \mn@doi [\mnras] {10.1093/mnras/stt601}, \href
  {http://adsabs.harvard.edu/abs/2013MNRAS.432.2433H} {432, 2433}

\bibitem[\protect\citeauthoryear{{Hildebrandt} et~al.,}{{Hildebrandt}
  et~al.}{2017}]{hendrick2017}
{Hildebrandt} H.,  et~al., 2017, \mn@doi [\mnras] {10.1093/mnras/stw2805},
  \href {http://adsabs.harvard.edu/abs/2017MNRAS.465.1454H} {465, 1454}

\bibitem[\protect\citeauthoryear{{Hirata} \& {Seljak}}{{Hirata} \&
  {Seljak}}{2004}]{hirata2004}
{Hirata} C.~M.,  {Seljak} U.,  2004, \mn@doi [\prd]
  {10.1103/PhysRevD.70.063526}, \href
  {http://adsabs.harvard.edu/abs/2004PhRvD..70f3526H} {70, 063526}

\bibitem[\protect\citeauthoryear{{Jee}, {Tyson}, {Hilbert}, {Schneider},
  {Schmidt}  \& {Wittman}}{{Jee} et~al.}{2016}]{jee2016}
{Jee} M.~J.,  {Tyson} J.~A.,  {Hilbert} S.,  {Schneider} M.~D.,  {Schmidt} S.,
   {Wittman} D.,  2016, \mn@doi [\apj] {10.3847/0004-637X/824/2/77}, \href
  {http://adsabs.harvard.edu/abs/2016ApJ...824...77J} {824, 77}

\bibitem[\protect\citeauthoryear{{Joachimi} \& {Bridle}}{{Joachimi} \&
  {Bridle}}{2010}]{joachimi2010}
{Joachimi} B.,  {Bridle} S.~L.,  2010, \mn@doi [\aap]
  {10.1051/0004-6361/200913657}, \href
  {http://adsabs.harvard.edu/abs/2010A%26A...523A...1J} {523, A1}

\bibitem[\protect\citeauthoryear{{Joachimi} \& {Schneider}}{{Joachimi} \&
  {Schneider}}{2009}]{joachimi2009}
{Joachimi} B.,  {Schneider} P.,  2009, \mn@doi [\aap]
  {10.1051/0004-6361/200912420}, \href
  {http://adsabs.harvard.edu/abs/2009A%26A...507..105J} {507, 105}

\bibitem[\protect\citeauthoryear{{Joachimi}, {Mandelbaum}, {Abdalla}  \&
  {Bridle}}{{Joachimi} et~al.}{2011}]{joachimi2011}
{Joachimi} B.,  {Mandelbaum} R.,  {Abdalla} F.~B.,   {Bridle} S.~L.,  2011,
  \mn@doi [\aap] {10.1051/0004-6361/201015621}, \href
  {http://adsabs.harvard.edu/abs/2011A%26A...527A..26J} {527, A26}

\bibitem[\protect\citeauthoryear{{Johnson} et~al.,}{{Johnson}
  et~al.}{2017}]{johnson2017}
{Johnson} A.,  et~al., 2017, \mn@doi [\mnras] {10.1093/mnras/stw3033}, \href
  {http://adsabs.harvard.edu/abs/2017MNRAS.465.4118J} {465, 4118}

\bibitem[\protect\citeauthoryear{{Joudaki} et~al.,}{{Joudaki}
  et~al.}{2017}]{joudaki2017}
{Joudaki} S.,  et~al., 2017, \mn@doi [\mnras] {10.1093/mnras/stx998}, \href
  {http://adsabs.harvard.edu/abs/2017MNRAS.471.1259J} {471, 1259}

\bibitem[\protect\citeauthoryear{{Joudaki} et~al.,}{{Joudaki}
  et~al.}{2018}]{joudaki2018}
{Joudaki} S.,  et~al., 2018, \mn@doi [\mnras] {10.1093/mnras/stx2820}, \href
  {http://adsabs.harvard.edu/abs/2018MNRAS.474.4894J} {474, 4894}

\bibitem[\protect\citeauthoryear{{Kirk} et~al.,}{{Kirk}
  et~al.}{2015}]{kirk2015}
{Kirk} D.,  et~al., 2015, \mn@doi [\ssr] {10.1007/s11214-015-0213-4}, \href
  {http://adsabs.harvard.edu/abs/2015SSRv..193..139K} {193, 139}

\bibitem[\protect\citeauthoryear{{Kitching}, {Miller}, {Heymans}, {van
  Waerbeke}  \& {Heavens}}{{Kitching} et~al.}{2008}]{lensfit2}
{Kitching} T.~D.,  {Miller} L.,  {Heymans} C.~E.,  {van Waerbeke} L.,
  {Heavens} A.~F.,  2008, \mn@doi [\mnras] {10.1111/j.1365-2966.2008.13628.x},
  \href {http://adsabs.harvard.edu/abs/2008MNRAS.390..149K} {390, 149}

\bibitem[\protect\citeauthoryear{{Kuijken}}{{Kuijken}}{2008}]{gaap}
{Kuijken} K.,  2008, \mn@doi [\aap] {10.1051/0004-6361:20066601}, \href
  {http://adsabs.harvard.edu/abs/2008A%26A...482.1053K} {482, 1053}

\bibitem[\protect\citeauthoryear{{Kuijken}}{{Kuijken}}{2011}]{omegacam}
{Kuijken} K.,  2011, The Messenger, \href
  {http://adsabs.harvard.edu/abs/2011Msngr.146....8K} {146, 8}

\bibitem[\protect\citeauthoryear{{Kuijken} et~al.,}{{Kuijken}
  et~al.}{2015}]{kuijken2015}
{Kuijken} K.,  et~al., 2015, \mn@doi [\mnras] {10.1093/mnras/stv2140}, \href
  {http://adsabs.harvard.edu/abs/2015MNRAS.454.3500K} {454, 3500}

\bibitem[\protect\citeauthoryear{{Leistedt} \& {Hogg}}{{Leistedt} \&
  {Hogg}}{2017}]{boris2017}
{Leistedt} B.,  {Hogg} D.~W.,  2017, \mn@doi [\apj] {10.3847/1538-4357/aa6332},
  \href {http://adsabs.harvard.edu/abs/2017ApJ...838....5L} {838, 5}

\bibitem[\protect\citeauthoryear{{Liske} et~al.,}{{Liske}
  et~al.}{2015}]{likse2015}
{Liske} J.,  et~al., 2015, \mn@doi [\mnras] {10.1093/mnras/stv1436}, \href
  {http://adsabs.harvard.edu/abs/2015MNRAS.452.2087L} {452, 2087}

\bibitem[\protect\citeauthoryear{{Mancone} \& {Gonzalez}}{{Mancone} \&
  {Gonzalez}}{2012a}]{ezgal_software}
{Mancone} C.,  {Gonzalez} A.,  2012a, {EzGal: A Flexible Interface for Stellar
  Population Synthesis Models}, Astrophysics Source Code Library (\mn@eprint
  {ascl} {1208.021})

\bibitem[\protect\citeauthoryear{{Mancone} \& {Gonzalez}}{{Mancone} \&
  {Gonzalez}}{2012b}]{ezgal_paper}
{Mancone} C.~L.,  {Gonzalez} A.~H.,  2012b, \mn@doi [\pasp] {10.1086/666502},
  \href {http://adsabs.harvard.edu/abs/2012PASP..124..606M} {124, 606}

\bibitem[\protect\citeauthoryear{{Mandelbaum} et~al.,}{{Mandelbaum}
  et~al.}{2005}]{mandelbaum2005}
{Mandelbaum} R.,  et~al., 2005, \mn@doi [\mnras]
  {10.1111/j.1365-2966.2005.09282.x}, \href
  {http://adsabs.harvard.edu/abs/2005MNRAS.361.1287M} {361, 1287}

\bibitem[\protect\citeauthoryear{{Mandelbaum} et~al.,}{{Mandelbaum}
  et~al.}{2011}]{mandelbaum2011}
{Mandelbaum} R.,  et~al., 2011, \mn@doi [\mnras]
  {10.1111/j.1365-2966.2010.17485.x}, \href
  {http://adsabs.harvard.edu/abs/2011MNRAS.410..844M} {410, 844}

\bibitem[\protect\citeauthoryear{{Mandelbaum}, {Slosar}, {Baldauf}, {Seljak},
  {Hirata}, {Nakajima}, {Reyes}  \& {Smith}}{{Mandelbaum}
  et~al.}{2013}]{mandelbaum2013}
{Mandelbaum} R.,  {Slosar} A.,  {Baldauf} T.,  {Seljak} U.,  {Hirata} C.~M.,
  {Nakajima} R.,  {Reyes} R.,   {Smith} R.~E.,  2013, \mn@doi [\mnras]
  {10.1093/mnras/stt572}, \href
  {http://adsabs.harvard.edu/abs/2013MNRAS.432.1544M} {432, 1544}

\bibitem[\protect\citeauthoryear{{Miller}, {Kitching}, {Heymans}, {Heavens}  \&
  {van Waerbeke}}{{Miller} et~al.}{2007}]{lensfit1}
{Miller} L.,  {Kitching} T.~D.,  {Heymans} C.,  {Heavens} A.~F.,   {van
  Waerbeke} L.,  2007, \mn@doi [\mnras] {10.1111/j.1365-2966.2007.12363.x},
  \href {http://adsabs.harvard.edu/abs/2007MNRAS.382..315M} {382, 315}

\bibitem[\protect\citeauthoryear{{Miller} et~al.,}{{Miller}
  et~al.}{2013}]{lensfit3}
{Miller} L.,  et~al., 2013, \mn@doi [\mnras] {10.1093/mnras/sts454}, \href
  {http://adsabs.harvard.edu/abs/2013MNRAS.429.2858M} {429, 2858}

\bibitem[\protect\citeauthoryear{{Morrison} \& {Hildebrandt}}{{Morrison} \&
  {Hildebrandt}}{2015}]{morrison2015}
{Morrison} C.~B.,  {Hildebrandt} H.,  2015, \mn@doi [\mnras]
  {10.1093/mnras/stv2103}, \href
  {http://adsabs.harvard.edu/abs/2015MNRAS.454.3121M} {454, 3121}

\bibitem[\protect\citeauthoryear{{Morrison}, {Hildebrandt}, {Schmidt},
  {Baldry}, {Bilicki}, {Choi}, {Erben}  \& {Schneider}}{{Morrison}
  et~al.}{2017}]{morrison2017}
{Morrison} C.~B.,  {Hildebrandt} H.,  {Schmidt} S.~J.,  {Baldry} I.~K.,
  {Bilicki} M.,  {Choi} A.,  {Erben} T.,   {Schneider} P.,  2017, \mn@doi
  [\mnras] {10.1093/mnras/stx342}, \href
  {http://adsabs.harvard.edu/abs/2017MNRAS.467.3576M} {467, 3576}

\bibitem[\protect\citeauthoryear{{Norberg}, {Baugh}, {Gazta{\~n}aga}  \&
  {Croton}}{{Norberg} et~al.}{2009}]{norberg2009}
{Norberg} P.,  {Baugh} C.~M.,  {Gazta{\~n}aga} E.,   {Croton} D.~J.,  2009,
  \mn@doi [\mnras] {10.1111/j.1365-2966.2009.14389.x}, \href
  {http://adsabs.harvard.edu/abs/2009MNRAS.396...19N} {396, 19}

\bibitem[\protect\citeauthoryear{{Prat} et~al.,}{{Prat}
  et~al.}{2017}]{prat2017}
{Prat} J.,  et~al., 2017, preprint, \href
  {http://adsabs.harvard.edu/abs/2017arXiv170801537P} {} (\mn@eprint {arXiv}
  {1708.01537})

\bibitem[\protect\citeauthoryear{{Radovich} et~al.,}{{Radovich}
  et~al.}{2017}]{radovich2017}
{Radovich} M.,  et~al., 2017, \mn@doi [\aap] {10.1051/0004-6361/201629353},
  \href {http://adsabs.harvard.edu/abs/2017A%26A...598A.107R} {598, A107}

\bibitem[\protect\citeauthoryear{{Rozo} et~al.,}{{Rozo}
  et~al.}{2016}]{rozo2016}
{Rozo} E.,  et~al., 2016, \mn@doi [\mnras] {10.1093/mnras/stw1281}, \href
  {http://adsabs.harvard.edu/abs/2016MNRAS.461.1431R} {461, 1431}

\bibitem[\protect\citeauthoryear{{Rykoff} et~al.,}{{Rykoff}
  et~al.}{2014}]{redmap_sdss}
{Rykoff} E.~S.,  et~al., 2014, \mn@doi [\apj] {10.1088/0004-637X/785/2/104},
  \href {http://adsabs.harvard.edu/abs/2014ApJ...785..104R} {785, 104}

\bibitem[\protect\citeauthoryear{{Rykoff} et~al.,}{{Rykoff}
  et~al.}{2016}]{redmap_des}
{Rykoff} E.~S.,  et~al., 2016, \mn@doi [\apjs] {10.3847/0067-0049/224/1/1},
  \href {http://adsabs.harvard.edu/abs/2016ApJS..224....1R} {224, 1}

\bibitem[\protect\citeauthoryear{{Sadeh}, {Abdalla}  \& {Lahav}}{{Sadeh}
  et~al.}{2016}]{sadeh2016}
{Sadeh} I.,  {Abdalla} F.~B.,   {Lahav} O.,  2016, \mn@doi [\pasp]
  {10.1088/1538-3873/128/968/104502}, \href
  {http://adsabs.harvard.edu/abs/2016PASP..128j4502S} {128, 104502}

\bibitem[\protect\citeauthoryear{{Schechter}}{{Schechter}}{1976}]{schecter1976}
{Schechter} P.,  1976, \mn@doi [\apj] {10.1086/154079}, \href
  {http://adsabs.harvard.edu/abs/1976ApJ...203..297S} {203, 297}

\bibitem[\protect\citeauthoryear{{Schlegel}, {Finkbeiner}  \&
  {Davis}}{{Schlegel} et~al.}{1998}]{schlegel98}
{Schlegel} D.~J.,  {Finkbeiner} D.~P.,   {Davis} M.,  1998, \mn@doi [\apj]
  {10.1086/305772}, \href {http://adsabs.harvard.edu/abs/1998ApJ...500..525S}
  {500, 525}

\bibitem[\protect\citeauthoryear{{Shirasaki}, {Takada}, {Miyatake},
  {Takahashi}, {Hamana}, {Nishimichi}  \& {Murata}}{{Shirasaki}
  et~al.}{2017}]{shirasaki2017}
{Shirasaki} M.,  {Takada} M.,  {Miyatake} H.,  {Takahashi} R.,  {Hamana} T.,
  {Nishimichi} T.,   {Murata} R.,  2017, \mn@doi [\mnras]
  {10.1093/mnras/stx1477}, \href
  {http://adsabs.harvard.edu/abs/2017MNRAS.470.3476S} {470, 3476}

\bibitem[\protect\citeauthoryear{{Singh}, {Mandelbaum}  \& {More}}{{Singh}
  et~al.}{2015}]{singh2015}
{Singh} S.,  {Mandelbaum} R.,   {More} S.,  2015, \mn@doi [\mnras]
  {10.1093/mnras/stv778}, \href
  {http://adsabs.harvard.edu/abs/2015MNRAS.450.2195S} {450, 2195}

\bibitem[\protect\citeauthoryear{{Singh}, {Mandelbaum}, {Seljak}, {Slosar}  \&
  {Vazquez Gonzalez}}{{Singh} et~al.}{2017}]{singh2017}
{Singh} S.,  {Mandelbaum} R.,  {Seljak} U.,  {Slosar} A.,   {Vazquez Gonzalez}
  J.,  2017, \mn@doi [\mnras] {10.1093/mnras/stx1828}, \href
  {http://adsabs.harvard.edu/abs/2017MNRAS.471.3827S} {471, 3827}

\bibitem[\protect\citeauthoryear{{Tonegawa}, {Okumura}, {Totani}, {Dalton}  \&
  {Yabe}}{{Tonegawa} et~al.}{2017}]{fastsound2017}
{Tonegawa} M.,  {Okumura} T.,  {Totani} T.,  {Dalton} G.,   {Yabe} K.,  2017,
  preprint, \href {http://adsabs.harvard.edu/abs/2017arXiv170802224T} {}
  (\mn@eprint {arXiv} {1708.02224})

\bibitem[\protect\citeauthoryear{{Troxel} et~al.,}{{Troxel}
  et~al.}{2017}]{troxel2017}
{Troxel} M.~A.,  et~al., 2017, preprint, \href
  {http://adsabs.harvard.edu/abs/2017arXiv170801538T} {} (\mn@eprint {arXiv}
  {1708.01538})

\bibitem[\protect\citeauthoryear{{VanderPlas}, {Fouesneau}  \&
  {Taylor}}{{VanderPlas} et~al.}{2014}]{astroml}
{VanderPlas} J.,  {Fouesneau} M.,   {Taylor} J.,  2014, {AstroML: Machine
  learning and data mining in astronomy}, Astrophysics Source Code Library
  (\mn@eprint {ascl} {1407.018})

\bibitem[\protect\citeauthoryear{{Viola} et~al.,}{{Viola}
  et~al.}{2015}]{viola2015}
{Viola} M.,  et~al., 2015, \mn@doi [\mnras] {10.1093/mnras/stv1447}, \href
  {http://adsabs.harvard.edu/abs/2015MNRAS.452.3529V} {452, 3529}

\bibitem[\protect\citeauthoryear{{Wadadekar}}{{Wadadekar}}{2005}]{Wadadekar2005}
{Wadadekar} Y.,  2005, \mn@doi [\pasp] {10.1086/427710}, \href
  {http://adsabs.harvard.edu/abs/2005PASP..117...79W} {117, 79}

\bibitem[\protect\citeauthoryear{{Way}, {Foster}, {Gazis}  \&
  {Srivastava}}{{Way} et~al.}{2009}]{Way2009}
{Way} M.~J.,  {Foster} L.~V.,  {Gazis} P.~R.,   {Srivastava} A.~N.,  2009,
  \mn@doi [\apj] {10.1088/0004-637X/706/1/623}, \href
  {http://adsabs.harvard.edu/abs/2009ApJ...706..623W} {706, 623}

\bibitem[\protect\citeauthoryear{{Wolf} et~al.,}{{Wolf}
  et~al.}{2017}]{wolf2017}
{Wolf} C.,  et~al., 2017, \mn@doi [\mnras] {10.1093/mnras/stw3151}, \href
  {http://adsabs.harvard.edu/abs/2017MNRAS.466.1582W} {466, 1582}

\bibitem[\protect\citeauthoryear{{York} et~al.,}{{York}
  et~al.}{2000}]{york2000}
{York} D.~G.,  et~al., 2000, \mn@doi [\aj] {10.1086/301513}, \href
  {http://adsabs.harvard.edu/abs/2000AJ....120.1579Y} {120, 1579}

\bibitem[\protect\citeauthoryear{{de Jong}, {Verdoes Kleijn}, {Kuijken}  \&
  {Valentijn}}{{de Jong} et~al.}{2013}]{kids}
{de Jong} J.~T.~A.,  {Verdoes Kleijn} G.~A.,  {Kuijken} K.~H.,   {Valentijn}
  E.~A.,  2013, \mn@doi [Experimental Astronomy] {10.1007/s10686-012-9306-1},
  \href {http://adsabs.harvard.edu/abs/2013ExA....35...25D} {35, 25}

\bibitem[\protect\citeauthoryear{{de Jong} et~al.,}{{de Jong}
  et~al.}{2017}]{kids_dr3}
{de Jong} J.~T.~A.,  et~al., 2017, \mn@doi [\aap]
  {10.1051/0004-6361/201730747}, \href
  {http://adsabs.harvard.edu/abs/2017A%26A...604A.134D} {604, A134}

\bibitem[\protect\citeauthoryear{{van Uitert} et~al.,}{{van Uitert}
  et~al.}{2016}]{edo2016}
{van Uitert} E.,  et~al., 2016, \mn@doi [\mnras] {10.1093/mnras/stw747}, \href
  {http://adsabs.harvard.edu/abs/2016MNRAS.459.3251V} {459, 3251}

\bibitem[\protect\citeauthoryear{{van Uitert} et~al.,}{{van Uitert}
  et~al.}{2018}]{edo2018}
{van Uitert} E.,  et~al., 2018, \mn@doi [\mnras] {10.1093/mnras/sty551}, \href
  {http://adsabs.harvard.edu/abs/2018MNRAS.tmp..592V} {}

\makeatother
\end{thebibliography}

%%%%%%%%%%%%%%%%%%%%%%%%%%%%%%%%%%%%%%%%%%%%%%%%%%

%%%%%%%%%%%%%%%%% APPENDICES %%%%%%%%%%%%%%%%%%%%%
%\clearpage

\appendix

\section{Distribution of the selected red galaxies in the colour-magnitude space}

\begin{figure*}
 \begin{tabular}{cc}
\includegraphics[width=\columnwidth]{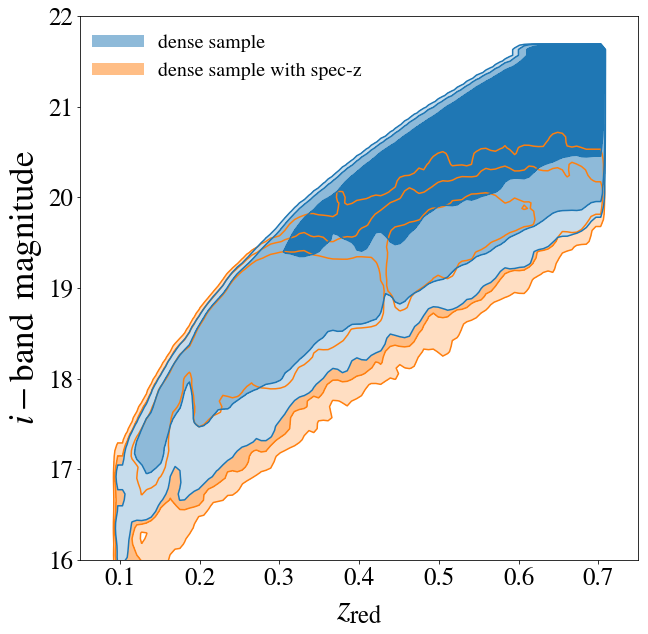}
\includegraphics[width=\columnwidth]{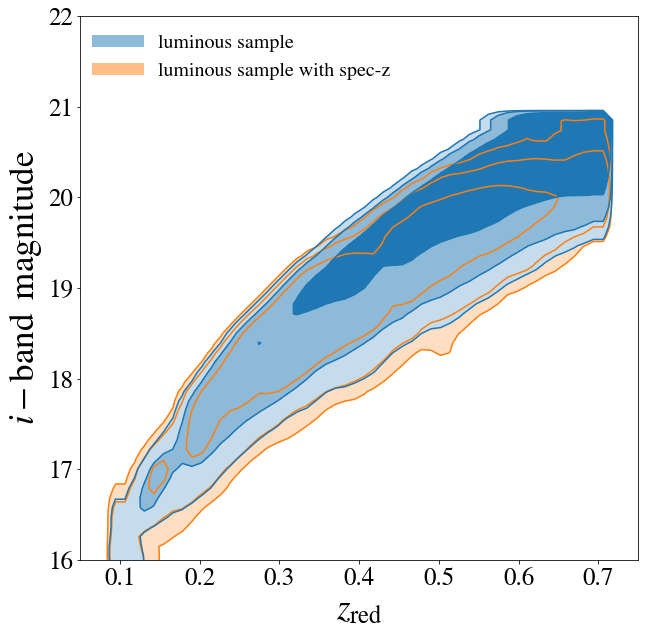}
\end{tabular}
\caption{\label{fig:miz} Distribution of LRGs in the two dimensional space of the $i$-band magnitude $m_i$ and red-sequence redshift $z_{\rm red}$. Shown are contours containing 68\%, 95\%, and 99\% of all selected LRGs (blue) and the selected LRGs with secure spectroscopic redshift (orange). The distribution of galaxies in the $\mathtt{dense}$ and in the $\mathtt{luminous}$ samples are shown in the Left and Right panels respectively. We note that in terms of magnitude distribution, galaxies in the $\mathtt{luminous}$ sample have a more representative sample of spectroscopic counterparts compared to galaxies in the $\mathtt{dense}$ sample.}
\end{figure*}

Compared to the galaxies with spectroscopic redshifts, the distribution of the selected galaxies in magnitude space is more inclined towards fainter galaxies. This is shown in Figure~\ref{fig:miz} for the $\dense$ sample (left panel) and the $\lum$ sample (right panel). In both samples, there are more galaxies at higher redshifts and higher $i$-band magnitudes. Figure~\ref{fig:miz} shows contours containing 68\%, 95\%, and 99\% of galaxy densities in the two dimensional space of $z_{\rm red}$ and $m_i$. The blue contours show the distribution of all selected LRGs while the orange contours show the distribution of LRGs with spectroscopy. The $z_{\rm red}-m_i$ distribution of galaxies in the $\mathtt{luminous}$ sample has a better match with the distribution of galaxies in the same sample that have spectroscopic redshifts. Since the galaxies in the $\lum$ sample are brighter, they have a more representative magnitude distribution in the spectroscopic data.

\begin{figure*}
 \begin{tabular}{cc}
\includegraphics[width=\columnwidth]{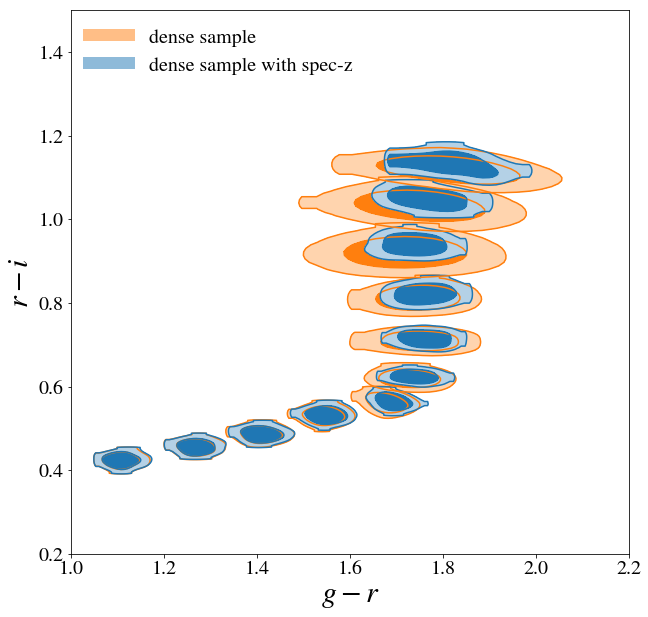}
\includegraphics[width=\columnwidth]{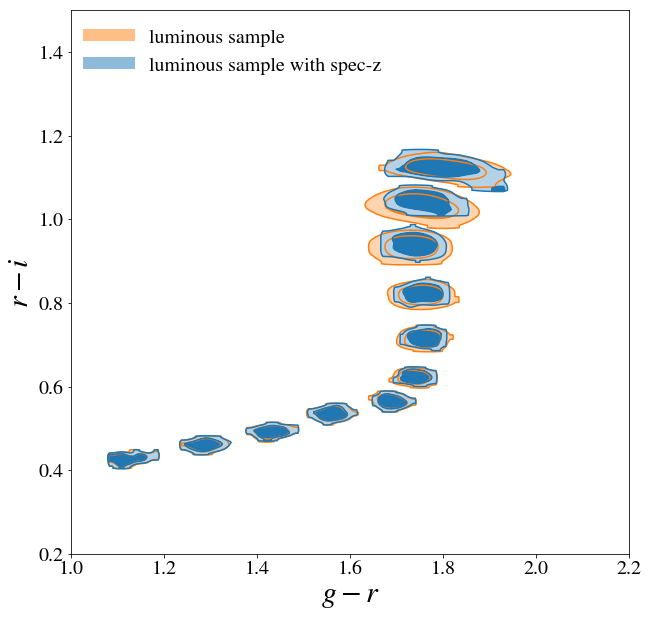}
\end{tabular}
\caption{\label{fig:colour_dense}Left:
68\% and 95\% of galaxy density contours in the $g-r$ and $r-i$ space for galaxies in the $\mathtt{dense}$ sample for different redshift bins. From bottom to top the redshift bins correspond to $z=0.16,0.21,0.26,0.31,0.36,0.41,0.46,0.51,0.56,0.61,0.67$ respectively. The blue contours correspond to galaxies in the $\mathtt{dense}$ sample that have spectroscopy, while the orange contours correspond to all galaxies in the $\dense$ sample with the exception that their maximum $m_i$ is set by the maximum $m_i$ of galaxies with spectroscopy in the same redshift bin. 
Right: same as the left panel but showing galaxies in the $\lum$ sample.
Any mismatch between the orange and the blue contours implies biased spectroscopic sampling of the selected LRGs. The spectroscopic sampling of both $\dense$ and $\lum$ galaxies is unbiased up to $z\simeq 0.36$ and mismatch between the colour distributions at higher redshift bins ($z \geq 0.4$) is evident. The biased spectroscopic sampling of the selected galaxies is slightly less pronounced in the $\lum$ sample.}
\end{figure*}

In the two dimensional space of $g-r$ and $r-i$ colours, we compare the redshift-dependent distribution of red galaxies with the distribution of red galaxies that have spectroscopy. These distributions are shown in Figure~\ref{fig:colour_dense} with the right (left) panel corresponding to galaxies in the $\dense$ ($\lum$) sample. In particular, we compute the 68\% and 95\% of galaxy densities in the $\{g-r, r-i\}$ colour space in bins of redshifts. The central values of the redshift bins are $\{0.16,0.21,0.26,0.31,0.36,0.41,0.46,0.51,0.56,0.61,0.67\}$ from bottom to top in each panel. The orange contours are the densities of the selected red galaxies while the blue contours show the densities of the red galaxies with spectroscopy.

We have seen in Figure~\ref{fig:miz} that the photometrically-selected red galaxies and those with spectroscopy do not sample the apparent magnitude-redshift space in a similar fashion. Therefore in each redshift bin in Figure~\ref{fig:colour_dense}, we apply a cut to the apparent magnitude of the selected red galaxies such that their maximum apparent magnitude is equal to the maximum apparent magnitude of the red galaxies with spectroscopy. This allows us to make a fair comparison between the two colour distributions. We note that at certain redshifts the colour distributions of all red galaxies and that of the red galaxies with spectroscopy do not fully overlap. In other words, at certain redshifts, spectroscopic data samples the colour space of the selected red galaxies in a biased manner which could lead to biases in the estimated photometric redshifts. For both $\dense$ and $\lum$ galaxies, this sampling bias becomes more pronounced for $z>0.36$. This leads to slight deviation of bias from zero at higher redshifts for both $\dense$ and $\lum$ galaxies. As shown in Figure~\ref{fig:colour_dense}, $\lum$ sample is a more representative sample of the spectroscopic dataset. Therefore, the median absolute value of redshift bias is slightly smaller for this sample. 
%Spectroscopic sampling of $\lum$ galaxies is more representative. As a result the estimated median absolute value of redshift bias is slightly lower for these galaxies.  

\bsp	% typesetting comment
\label{lastpage}
\end{document}